\documentclass[12pt]{article}
\usepackage{graphicx}
\usepackage{epsfig}
\usepackage[square,comma,sort&compress]{natbib}
\renewcommand{\cite}{\citep}

\newcommand {\ignore}[1]{}

\def\nn{\nonumber}

\def\be{\begin{equation}}
\def\ee{\end{equation}}
\def\bear{\be\begin{array}}
\def\eear{\end{array}\ee} 
\def\bea{\begin{eqnarray}} 
\def\eea{\end{eqnarray}}

\def\vb#1{\vbox to #1 pt{}}
\def\beqa{\begin{eqnarray}}
\def\eeqa{\end{eqnarray}}

\def\beq{\begin{equation}}
\def\eeq{\end{equation}}
\def\ba{\begin{array}}
\def\ea{\end{array}}

\def\npb#1#2#3{{\it Nucl.\ Phys.\ }{\bf B #1} (#2) #3}
\def\plb#1#2#3{{\it Phys.\ Lett.\ }{\bf B #1} (#2) #3}
\def\prd#1#2#3{{\it Phys.\ Rev.\ }{\bf D #1} (#2) #3}

\def\prl#1#2#3{{\it Phys.\ Rev.\ Lett.\ }{\bf #1} (#2) #3}

\def\hepph#1{{\tt hep-ph/#1}}

\def\21{$SU(2) \otimes U(1)$}

\def\half{{\textstyle{1 \over 2}}} 
 
\def\quarter{{\textstyle{1 \over 4}}} 
 
\def\eighth{{\textstyle{1 \over 8}}} 
 
\def\bold#1{\setbox0=\hbox{$#1$} 
     \kern-.025em\copy0\kern-\wd0 
     \kern.05em\copy0\kern-\wd0 
     \kern-.025em\raise.0433em\box0 } 
\begin{document}

\begin{titlepage}

{\par\raggedleft hep-ph/0202054 \\
\par\raggedleft FISIST/02--2002/CFIF\\}

\vspace*{3mm}

{\par\centering \textbf{\large Charged Lepton Flavor Violation in Supersymmetry
with Bilinear R--Parity Violation}\large \par}

{\par\centering D. F. Carvalho, M. E. G\'{o}mez and J. C. Rom\~{a}o  \vspace{0.3cm}\\
\par}

{\par\centering \textit{Departamento de F\'{\i}sica, Instituto Superior T\'{e}cnico}\\
 \textit{Av. Rovisco Pais 1, \( \: \:  \) 1049-001 Lisboa, Portugal }\\
\par}

\begin{abstract}

The simplest unified extension of the Minimal Supersymmetric Standard
Model with bi-linear R--parity violation naturally predicts a
hierarchical neutrino mass spectrum, suitable to explain atmospheric
and solar neutrino fluxes.  We study whether the individual violation
of the lepton numbers $L_{e,\mu,\tau}$ in the charged sector can lead
to measurable rates for $BR(\mu\rightarrow e \gamma)$ and
$BR(\tau\rightarrow \mu \gamma)$.  We find that some of the R--parity
violating terms that are compatible with the observed atmospheric
neutrino oscillations could lead to rates for $\mu\rightarrow e
\gamma$ measurable in projected experiments. However, the $\Delta
m^2_{12}$ obtained for those parameters is too high to be compatible
with the solar neutrino data, excluding therefore the possibility of
having measurable rates for $\mu \rightarrow e \gamma$ in the model.

\end{abstract}

{\small
\vspace{90mm}
\line(1,0){140}

Email addresses: dani@cfif.ist.utl.pt,
mgomez@cfif.ist.utl.pt,
jorge.romao@ist.utl.pt
}

\end{titlepage}

\newpage

\setcounter{page}{1}

\section{Introduction}

In the Standard Model (SM), lepton number is exactly preserved in
contradiction with the observed neutrino oscillations
\cite{kamiokande,chooz}. Several extension of the SM include patterns of
neutrino masses and mixings which can provide a satisfactory
explanation for these flavor oscillations.  The consequences of the
individual violation of the lepton numbers $L_{e,\mu,\tau}$ for
charged lepton will be manifest in processes such as $\mu\rightarrow e
\gamma$, $\mu\rightarrow 3 e$, $\mu-e$ conversion in heavy nuclei,
$\tau\rightarrow \mu \gamma$ and $K_L\rightarrow \mu e$ \cite{kaon}. The
experimental upper bound for these processes is quite restrictive,
which imposes a significant constraint for the explanation of flavor in
models beyond the SM. However, the mechanisms used to explain the
origin of the tiny neutrino masses required to explain solar and
atmospheric neutrino oscillations, typically imply that these processes
may occur at small rates, motivating an increasing experimental
interest in exploring further charged lepton flavor violating processes.

The rates for charged lepton flavor violation (LFV) are extremely
small in the SM with right--handed neutrinos ($\propto \Delta
m_\nu^4/M_W^4$ \cite{P..}).  In R--parity conserving supersymmetric
(SUSY) models, like the Minimal Supersymmetric Standard Model (MSSM),
the presence of LFV processes is associated with vertices involving
leptons and their superpartners \cite{rh-nu}. These processes are
sensitive to the scalar mass matrices structure, a non--diagonality of
the latter in a basis in which fermions are diagonal, leads to a hard
violation of flavor. The structure of the scalar mass matrices is very
sensitive to the SUSY--breaking, in particular in models where SUSY is
softly broken, LFV imposes a severe constraint in the flavor
dependence of the soft--terms as they are generated in GUT's and
string inspired models \cite{Carvalho:2000xg}.

The inclusion of a ``see--saw'' mechanism in the MSSM provides a very
attractive scenario to understand neutrino oscillations with very
small neutrino masses, and at the same time gives rates for LFV processes
accessible in projected experiments \cite{hisano,Carvalho:2001ex}.
The waiving of the R-parity symmetry in the MSSM provides an
alternative scenario to explain the generation of small neutrino
masses. In this case the R-parity violating operators can be
constrained by rare processes 
\cite{frank,r-par,Cheung:2001sb,Hirsch:1999kc}.

The simplest extension of the MSSM with bilinear R--parity violation
(BRpV) \cite{e3others} (allowing B--conserving but L-violating
interactions) can explain neutrino masses and mixings which can
account for the observed neutrino oscillations \cite{Romao:1999up}.  
The BRpV model has been extensively discussed in the
literature~\cite{epsrad}. It is motivated  by the fact
that it provides an effective truncation of models where R--parity
breaks \emph{spontaneously} by singlet sneutrino vev's around the weak
scale~\cite{SBRpV}. Moreover, they allow for the radiative breaking of
R-parity, opening also new ways to unify Gauge and Yukawa
couplings~\cite{Diaz:1999wz} and with a potentially slightly lower
prediction for \( \alpha _{s} \)~\cite{Diaz:1999is}. For recent papers
on phenomenological implications of these models see
Ref.~\cite{rphen00,chargedhiggs}.  As the parameters
involved in the R--parity violating operator are  constrained in order to
predict neutrino masses in the sub-eV range, we address in this paper 
the question of whether this operator will induce rates for charged
LFV processes of experimental interest. Some of them occur at
tree--level such as double $\beta$ decay \cite{dbd,Hirsch:1999kc} and $\mu-e$
conversion in nuclei \cite{Faessler:pn}. One loop
LFV decays as $l_j \rightarrow l_i \gamma$ become interesting on this
framework due to the experimental interest in improving the current
limits \cite{lfv-limit}:
\begin{eqnarray}
BR(\mu \to e \gamma) &<& 1.2 \times 10^{-11} \nn\\ BR(\tau \to \mu
\gamma) &<& 1.1 \times 10^{-6} \nn\\ BR(\tau \to e \gamma) &<& 2.7
\times 10^{-6}.
\label{limit}
\end{eqnarray}
As we will show, the predictions for the last two processes are much
lower than the above limits and will not constrain the BRpV model.
For $\mu \rightarrow e \gamma$ the predictions are compatible with the
current limit but could begin to constrain the model for the bounds
that will be reached in current \cite{psi} or planned experiments
\cite{prism}, if only the atmospheric neutrino data were taken in
account.  However the requirement that the one--loop induced $\Delta
m^2_{12}$ is in agreement with the solar neutrino data will imply that
the predicted rates for $\mu \rightarrow e \gamma$ will not be
visible, even in those new experiments.

This paper is organized as follows. In sections \ref{sec:TheModel},
\ref{sec:ThePot} and \ref{sec:MassMatrices} we describe the model, the
scalar potential and the fermion mass matrices, respectively. In
section~\ref{sec:Expressions} we derive the expressions for the LFV
processes. The results are presented in section~\ref{sec:Results} and
in section~\ref{sec:Conclusions} we give our conclusions.  The more
technical questions regarding the mass matrices, couplings and the
explicit formulas for the amplitudes are given in the appendices.

\section{The Superpotential and the Soft Breaking Terms}

\label{sec:TheModel}

Using the conventions of Refs.~\cite{chargedhiggs,HabKaneGun} we introduce
the model by specifying the superpotential, which includes BRpV \cite{epsrad}
in three generations. It is given by 
\begin{equation}
\label{eq:Wsuppot}
W=\varepsilon _{ab}\left[
h_{U}^{ij}\widehat{Q}_{i}^{a}\widehat{U}_{j}\widehat{H}_{u}^{b} +
h_{D}^{ij}\widehat{Q}_{i}^{b}\widehat{D}_{j}\widehat{H}_{d}^{a} +
h_{E}^{ij}\widehat{L}_{i}^{b}\widehat{R}_{j}\widehat{H}_{d}^{a}-\mu
\widehat{H}_{d}^{a}\widehat{H}_{u}^{b}+\epsilon
_{i}\widehat{L}_{i}^{a}\widehat{H}_{u}^{b}\right]
\end{equation}
where the couplings \( h_{U} \), \( h_{D} \) and \( h_{E} \) are \(
3\times 3 \) Yukawa matrices and \( \mu \) and \( \epsilon _{i} \) are
parameters with units of mass. The second bilinear term in
Eq.~(\ref{eq:Wsuppot}) violates lepton number and therefore also
breaks R--parity.  The inclusion of the R--parity violating term, though
small, can modify the predictions of the MSSM. The most salient
features are that neutrinos become massive and the lightest neutralino
is no longer a dark matter candidate because it is 
allowed to decay. Furthermore, we
can observe that this model implies the mixing of the leptons with the
usual charginos and neutralinos of the MSSM. Lepton Yukawa couplings can be
written as diagonal matrices without any loss of generality since it
is possible to rotate the superfields \( {\hat{L}^{b}}_{i} \) in the
superpotential, Eq.~(\ref{eq:Wsuppot}), such that Yukawa matrix \(
h_{E} \) becomes diagonal. Conversely, in BRpV models it is possible
to apply a similar rotation to reduce the number \( \epsilon \)
parameters and provide a non-trivial structure to \( {h}_{E} \)
\cite{Gomez:1998sd}.

Supersymmetry breaking is parameterized with a set of soft supersymmetry 
breaking terms. In the MSSM these are given by 
\begin{equation}
  \label{eq:lsoft}
  {\cal L}_{soft}=- V_{soft}^{MSSM} + \left[
 \half M_{s}\lambda _{s}\lambda _{s}+\half M\lambda \lambda +\half
 M'\lambda' \lambda' +h.c.\right]
\end{equation}
where
\begin{eqnarray}
V_{soft}^{MSSM} & = &
 M_{Q}^{ij2}\widetilde{Q}^{a*}_{i}\widetilde{Q}^{a}_{j} +
 M_{U}^{ij2}\widetilde{U}_{i}\widetilde{U}^{*}_{j}
 +M_{D}^{ij2}\widetilde{D}_{i}\widetilde{D}^{*}_{j}+
 M_{L}^{ij2}\widetilde{L}^{a*}_{i}\widetilde{L}^{a}_{j}+
 M_{R}^{ij2}\widetilde{R}_{i}\widetilde{R}^{*}_{j}\nonumber \\ & & \!
 \! \! \!
 +m_{H_{d}}^{2}H^{a*}_{d}H^{a}_{d}+m_{H_{u}}^{2}H^{a*}_{u}H^{a}_{u} 
\label{eq:Vsoft} \\ & & \! \! \! \!
 +\varepsilon _{ab}\left[
 A_{U}^{ij}\widetilde{Q}_{i}^{a}\widetilde{U}_{j}H_{u}^{b} +
 A_{D}^{ij}\widetilde{Q}_{i}^{b}\widetilde{D}_{j}H_{d}^{a}+
 A_{E}^{ij}\widetilde{L}_{i}^{b}\widetilde{R}_{j}H_{d}^{a} - B\mu
 H_{d}^{a}H_{u}^{b}\right] \, .\nonumber
\end{eqnarray}
 In addition to the MSSM soft SUSY breaking terms in \(
V_{soft}^{MSSM} \) the BRpV model contains the following
extra term
\begin{equation}
\label{softBRpV}
V_{soft}^{BRpV}=-B_{i}\epsilon _{i}\varepsilon
_{ab}\widetilde{L}_{i}^{a}H_{u}^{b}\, ,
\end{equation}
where the \( B_{i} \) have units of mass.

The electroweak symmetry is broken when the two Higgs doublets \( H_{d} \)
and \( H_{u} \), and the neutral component of the slepton doublets 
\( \widetilde{L}_{i} \) acquire vacuum expectation values. We
introduce the notation: 
\begin{equation}
\label{eq:shiftdoub}
H_{d}={{H^{0}_{d}}\choose {H^{-}_{d}}}\, ,\qquad H_{u} =
{{H^{+}_{u}}\choose {H^{0}_{u}}}\, ,\qquad \widetilde{L}_{i}=
{{\tilde{L}^{0}_{i}}\choose {\tilde{\ell }^{-}_{i}}}\, ,
\end{equation}
where we shift the neutral fields with non--zero vev's as
\begin{equation}
\label{shiftedfields}
H_{d}^{0}\equiv {1\over {\sqrt{2}}}[\sigma ^{0}_{d} + v_{d}+i\varphi
^{0}_{d}]\, ,\quad H_{u}^{0}\equiv {1\over {\sqrt{2}}}[\sigma ^{0}_{u}
+v_{u}+i\varphi ^{0}_{u}]\, ,\quad \tilde{L}_{i}^{0}\equiv {1\over
{\sqrt{2}}}[\tilde{\nu }^{R}_{i}+v_{i}+i\tilde{\nu }^{I}_{i}]\, .
\end{equation}
Note that the \( W \) boson acquires a mass 
\( m_{W}^{2}=\quarter g^{2}v^{2} \), where \( v^{2}\equiv v_{d}^{2}+
v_{u}^{2}+v_{1}^{2}+v_{2}^{2}+v_{3}^{2}\simeq (246\; {\textrm{GeV})^{2}} \).

In addition to the above MSSM parameters, our model
contains nine new parameters, \( \epsilon _{i} \), \( v_{i} \) and \( B_{i} \).
The minimization of the scalar potential allows to relate some of these free
parameters. The values of \( \epsilon _{i} \), \( v_{i} \) are directly related
with the neutrino masses and mixings as we will discuss below.

\section{The Scalar Potential}

\label{sec:ThePot}

The electroweak symmetry is broken when the neutral Higgses and the
neutral slepton fields acquire non--zero vev's. These are calculated
via the minimization of the effective potential or, in the
diagrammatic method, via the tadpole equations. The full scalar
potential at tree level is
\begin{equation}
\label{V}
V_{total}^{0}=\sum _{i}\left| {\partial W\over \partial z_{i}}\right|
^{2}+V_{D}+V_{soft}^{MSSM}+V_{soft}^{BRpV}
\end{equation}
where \( z_{i} \) is any one of the scalar components of the
superfields in the superpotential in Eq.~(\ref{eq:Wsuppot}), \( V_{D}
\) are the \( D \)-terms, and \( V_{soft}^{BRpV} \) is given in
Eq.~(\ref{softBRpV}).

\noindent
The tree level scalar potential contains the following linear terms 
\begin{equation}
\label{eq:Vlinear}
V_{linear}^{0}=t_{d}^{0}\sigma ^{0}_{d}+t_{u}^{0}\sigma^{0}_{u} +
t_{1}^{0}\tilde{\nu }^{R}_{1} + t_{2}^{0}\tilde{\nu }^{R}_{2} +
t_{3}^{0}\tilde{\nu }^{R}_{3}\, ,
\end{equation}
where the different \( t^{0} \) are the tadpoles at tree level, their
explicit expressions can be found in Ref. \cite{Romao:1999up}.
The five tree level tadpoles \( t_{\alpha }^{0} \) are equal to zero at the
minimum of the tree level potential, therefore we can use them to 
express the parameters:
\begin{equation}
\mu ,\  B,\ B_{1}\ , B_{2}, \ B_{3},
\end{equation}
in terms of: 
\begin{equation}
v_{u},v_{d},\  \epsilon _{i},\  v_{i},\ M^{2}_{Lij},\ m^{2}_{H_{u}},\ m^{2}_{H_{d}}.
\end{equation}

\noindent
We have two possible solutions for $\mu$:
\begin{equation}
\label{mueq}
\mu =\frac{-b\pm \sqrt{b^{2}-4ac}}{2a}
\end{equation}
where $a=v^{2}_{u}-v^{2}_{d}$, $b=2v_{d}\sum ^{3}_{i=1}\epsilon _{i}v_{i}$ and
\begin{equation}
c=v^{2}_{u}\left(\sum ^{3}_{i=1}\epsilon ^{2}_{i} +
m^{2}_{H_{u}}\right)-v^{2}_{d}m^{2}_{H_{d}}-
\left(\sum ^{3}_{i=1}\epsilon _{i}v_{i}\right)^{2}-D v^{2}
-\frac{1}{2}\sum ^{3}_{i=1}\sum^{3}_{j=1}v_{i}v_{j}(M^{2}_{Lij}
+M^{2}_{Lji})
\end{equation}
where we have defined
$D=\eighth(g^2+g'^2)(v_1^2+v_2^2+v_3^2+v_d^2-v_u^2)$.

As one can easily verify, the above relations lead to the MSSM 
relation for $\mu^2$ in the limit of vanishing $\epsilon_i$ and $v_i$. The 
uncertainty of the sign in the MSSM is translated here into two possible 
values for $\mu$--term. However for the values of  $\epsilon_i$ and $v_i$ 
relevant to our work both solutions are close in module and of opposite 
sign.
The values for \( B \) and $B_i$~'s can be expressed  in terms of $\mu$ as:
\begin{eqnarray}
B&=&\frac{1}{v_{u}}\left[\frac{v_{d}}{\mu }(m_{H_{d}}^{2}+
\mu ^{2}+D)-\sum ^{3}_{i=1}\epsilon _{i}v_{i}\right]\\
B_{i}&=&\frac{1}{v_{u}}\left[v_{d}\mu - 
\sum ^{3}_{j=1}\epsilon _{j}v_{j}-\frac{1}{\epsilon _{i}}(Dv_{i}+
\frac{1}{2}\sum ^{3}_{j=1}v_{j}(M^{2}_{Lij}+M^{2}_{Lji}))\right]
\end{eqnarray}

The equivalent equations for the MSSM equations are obtained by setting 
\( \epsilon _{i},v_{i} \) equal to zero.

\section{ Fermion Masses with BRpV}

\label{sec:MassMatrices}

As we discussed in the previous section the presence of BRpV terms in
the superpotential, Eq.~(\ref{eq:Wsuppot}), induces non--zero vev's
for the sneutrinos and enables the neutrinos to have a mass, with a
value related to the size of \( \epsilon _{i},\, \, v_{i} \) and the
SUSY parameters involved on the electroweak symmetry
breaking. Furthermore, the non--conservation of the R--parity allows
the SUSY partners to mix with the SM particles. In this section we
describe with detail the resulting neutralino--neutrino and
chargino--charged--lepton mass matrices, since they are the most
directly related to our problem. The complete set of mass matrices for
the BRpV Model can be found in Ref.~\cite{Romao:1999up}.

\subsection{Neutralino--Neutrino Mass Matrix}

The range of values of the $\epsilon$--parameters is indirectly
associated to the size of the neutrino masses predicted by the
model. To explore this relation we describe next the mass mixings
among neutralinos and neutrinos.  In the basis \( \psi ^{0T}=
(-i\lambda' ,-i\lambda ^{3},\widetilde{H}_{d}^{0},
\widetilde{H}_{u}^{0},\nu _{e},\nu _{\mu },\nu _{\tau }) \) the
neutral fermion mass matrix \( {\bold M}_{N} \) is given by
\beq
{\bold M}_N=\left[ \begin{array}{cc}  
{\mathcal{M}}_{\chi^{0}}& m^{T} \cr
\vb{20}
m & 0 \cr
\end{array} \right] 
\eeq
where,
\beq
{\mathcal{M}}_{\chi^{0}}\hskip -2pt=\hskip -4pt 
\left[ \hskip -7pt \begin{array}{cccc}  
M_{1} & 0 & -\frac {1}{2}g^{\prime }v_{d} & \frac {1}{2}g^{\prime }v_{u} \cr
\vb{12}   
0 & M_{2} & \frac {1}{2}gv_{d} & -\frac {1}{2}gv_{u} \cr
\vb{12}   
-\frac {1}{2}g^{\prime }v_{d} & \frac {1}{2}gv_{d} & 0 & -\mu  \cr
\vb{12}
\frac {1}{2}g^{\prime }v_{u} & -\frac {1}{2}gv_{u} & -\mu & 0  \cr
\end{array} \hskip -6pt \right] 
\eeq
\noindent is the standard MSSM neutralino mass matrix and 
\beq
m=\left[ \begin{array}{cccc}  
-\frac {1}{2}g^{\prime }v_{1} & \frac {1}{2}gv_{1} & 0 & \epsilon _{1} \cr
\vb{18}
-\frac {1}{2}g^{\prime }v_{2} & \frac {1}{2}gv_{2} & 0 & \epsilon _{2}  \cr
\vb{18}
-\frac {1}{2}g^{\prime }v_{3} & \frac {1}{2}gv_{3} & 0 & \epsilon _{3}  \cr  
\end{array} \right] 
\eeq
characterizes the breaking of R-parity. 

The mass matrix \( {\bold M}_{N} \) is diagonalized by
\beq
{\mathcal{N}}^{*}\bold M_N{\mathcal{N}}^{-1}=diag(m_{\chi^{0}_i},
m_{\nu_j}),  
\label{chi0massmat} 
\eeq
where \( (i=1,\cdots ,4) \) for the neutralinos, and \( (j=1,\cdots ,3) \)
for the neutrinos.

Since \( m_{\nu }\ll m_{\chi ^{0}} \) the mass matrix \( {\bold M}_{N} \) is
similar to the \textit{''see--saw''} mass matrices and takes approximately
the form diag(\( {\mathcal{M}}_{\chi ^{0}},m_{eff} \)), with:
\beq
\label{eq:meff}
m_{eff} = - m \, {\mathcal{M}}_{\chi ^{0}}^{-1}\, m^{T} 
= \frac{M_{1} g^{2} \!+\! M_{2} {g'}^{2}}{4\, 
\det({\mathcal{M}}_{\chi^{0}})} \left(\hskip -2mm
\begin{array}{ccc}
\Lambda_{e}^{2} 
\hskip -1pt&\hskip -1pt
\Lambda_{e} \Lambda_{\mu
}\hskip -1pt&\hskip -1pt
\Lambda_{e} \Lambda_{\tau }\\
\Lambda_{e} \Lambda_{\mu 
}\hskip -1pt&\hskip -1pt
\Lambda_{\mu}^{2}
\hskip -1pt&\hskip -1pt
\Lambda_{\mu }\Lambda_{\tau }\\
\Lambda_{e} \Lambda_{\tau 
}\hskip -1pt&\hskip -1pt 
\Lambda_{\mu }\Lambda_{\tau 
}\hskip -1pt&\hskip -1pt
\Lambda_{\tau}^{2}
\end{array}\hskip -3mm \right). 
\eeq
where the $ \Lambda_i$ parameters in Eq.~(\ref{eq:meff}) are defined as: 
\beq
\Lambda_i \equiv \mu v_i + v_d \epsilon_i
\label{lambdai} 
\eeq
One of the neutrino species acquire a tree level non--zero mass, given by:
\beq
\label{mnutree} m_{\nu_{3}} = Tr(m_{eff}) = 
\frac{M_{1} g^{2} + M_{2} {g'}^{2}}{4\, \det({\mathcal{M}}_{\chi^{0}})}
|\vec {\Lambda}|^{2}, 
\eeq
where $|\vec {\Lambda}|^{2}\equiv \sum_{i=1}^3 \Lambda_i^2$. 
The two other neutrinos can get masses at one--loop as 
it is discussed in Ref.~\cite{Romao:1999up}. For our purposes it will
be important to have an estimate of the values of $\Delta m^2_{12}=
m^2_{\nu_2}- m^2_{\nu_1}$. We will use the results of
Ref.~\cite{Corfu2001} where it was found that, to a very good
approximation, $m_{\nu_1}=0$ and 
\begin{equation}
  \label{eq:mneu2}
  m_{\nu_2}=\frac{3}{16\pi^2}\, m_b \, \sin 2 \theta_b
  \,\frac{h^2_b}{\mu^2}\, \log \frac{m^2_{\tilde{b}_2}}{m^2_{\tilde{b}_1}}\
  \frac{\left( \vec \epsilon \times \vec \Lambda \right)^2 }{|\vec
  \Lambda |^2}
\end{equation}

The explanation of the data on neutrino oscillations given in
Ref.~\cite{Romao:1999up} requires the neutrino masses to be be on the
sub-eV range in order to fit the data on atmospheric neutrino
oscillations. In our examples we take a $m_{\nu_3} = 0.1\ \rm{eV}$,
which leads to values of the $|\vec {\Lambda}|$ in the range of
$0.1-1\ \rm{GeV}^2$, for the values of the SUSY parameters that we
will consider. Considering that we take positive values for $\mu$ we
should also take negative values for the product $\epsilon_i v_i$ to
avoid our analysis to be constrained to small values of $\epsilon_i$.
However, as we will see, for the values of the SUSY parameters that
give the largest $BR(\mu \rightarrow e \gamma)$, the values of the
$|\epsilon_i|$ have to be below 0.1 GeV, if $\Delta m^2_{12} <
10^{-4}$ (eV)${}^2$, as required by the solar neutrino data.

\subsection{Chargino--Charged Lepton Mass Matrix}

Due to the R-parity violating terms in the superpotential,
Eq.~(\ref{eq:Wsuppot}), the charginos mix with the charged leptons,
linking therefore the problem of the masses of the neutrinos with the
problem of the charged lepton flavor violation. We describe in this
subsection the chargino--lepton mass matrix to explain how the flavor
mixing on the charged lepton sector arises.
In a basis where  
\begin{equation}
\label{eq:basisch}
\psi ^{+T}=(-i\lambda ^{+},
\widetilde{H}_{u}^{+},e_{R}^{+},\mu _{R}^{+},\tau _{R}^{+}) \qquad 
{\rm{and}} \qquad 
\psi ^{-T}=(-i\lambda ^{-},\widetilde{H}_{d}^{-},e_{L}^{-},
\mu _{L}^{-},\tau _{L}^{-}) 
\end{equation}
the corresponding charged fermion mass terms in the Lagrangian are 
\begin{equation}
\label{eq:chFmterm}
{\mathcal{L}}_{m}=-{1\over 2}(\psi ^{+T},\psi ^{-T})\left( \begin{array}{cc}
0 & M^{T}_{C}\\
M_{C} & 0
\end{array}\right) \left( \matrix {\psi ^{+}\cr \psi ^{-}}\right) +h.c.
\end{equation}
where the chargino--charged lepton mass matrix $M_C$  is given in the Appendix
\ref{ap:A}.
As in the MSSM, $M_C$ is diagonalized by two rotation matrices, and we
include the physical charged leptons and charginos into a set 
of five charged fermions defined as:
\beq
F_i^{-}=U_{ij}\, \psi_j^{-} \quad ; \quad F_i^{+}=V_{ij}\, \psi_j^{+} 
\eeq
Such that 
\beq
\bold{U}^{*} \bold{M_{C}} \bold{V}^{-1} = \bold{M_{CD}} 
\eeq
where \( \bold {M_{CD}} \) is the diagonal charged fermion mass matrix.

In the previous expressions the \( F_{i}^{\pm } \) are two component spinors.
We construct the four component Dirac spinors out of the two component spinors
with the conventions\footnote{%
Here we depart from the conventions of Ref.~\cite{HabKaneGun} because we want
the \( e^{-} \), \( \mu ^{-} \) and \( \tau ^{-} \) to be the particles and
not the anti--particles.
}, 
\beq
\chi_i^{-}=\left(\matrix{
F_{i}^{-} \cr
\vb{18}
\overline{F_{i}^{+}}\cr} \right) 
\eeq

The parameterization of the matrices $V$, and $U$ given in
Appendix~\ref{ap:A}, that was introduced in
Ref.~\cite{dbd,original_ref}, provides a very accurate representation
of the exact result. By comparing the numerical results with the
analytical expressions shown in Appendix A, we found discrepancies of
less than the $1\%$. To obtain this level of accuracy we had to
introduce corrections in the definition of $V_L$, $V_R$, and
$\Omega_R$ with respect to the formulas of Ref.~\cite{dbd}. These
arise mainly from including the sub-matrix $E^\prime$ in our
derivation (see Appendix A). Although the size of the matrix 
elements of $E^\prime$ is
smaller than the other components of $M_C$, it must be taken into
account in order to match the results of the smaller elements of $U$
and $V$ found in the exact diagonalization.  Our definition of $V_L$
and $V_R$ leads to the correct form for the lower right $3 \times 3$
sub-matrices of $U$ and $V$. We will make use of it to explain the
details of our results. The inclusion of the matrix $E^\prime$ in the
determination of $\Omega_R$ allows to display the dependence of the
matrix elements on the $\Lambda$-parameters, rather than a explicit
dependence on the $\epsilon$'s as quoted in Ref.~\cite{dbd}.

Also, we must observe that the elements of $\Omega_L$ exceed the ones
of $\Omega_R$ by several orders of magnitude. Therefore we can
anticipate that the couplings containing the matrix $V$ in
Appendix~\ref{ap:C}, will be suppressed with respect to the ones
containing the elements of $U$.

\section{$l_j\rightarrow l_i \gamma$ Flavor Violating Processes and 
the $\mu$ Ano\-malous Magnetic Moment}

\label{sec:Expressions}

\subsection{Effective Lagrangian and Diagrams}

The effective operators that generate the decays  
$ l^{-}_{j}\rightarrow l^{-}_{i}\gamma$ and the lepton anomalous magnetic 
moment can be written as:
\begin{equation}
\label{leffg2}
\mathcal{L}_{eff}=e\, \frac{m_{l_{j}}}{2}\, \overline{l}_{i}
\sigma _{\mu \nu } F^{\mu \nu} \left(A_{Lij}P_{L}+A_{Rij}P_{R}\right)l_{j}
\end{equation}

\begin{figure}[htbp]
\includegraphics[bb=0 15 440 155]{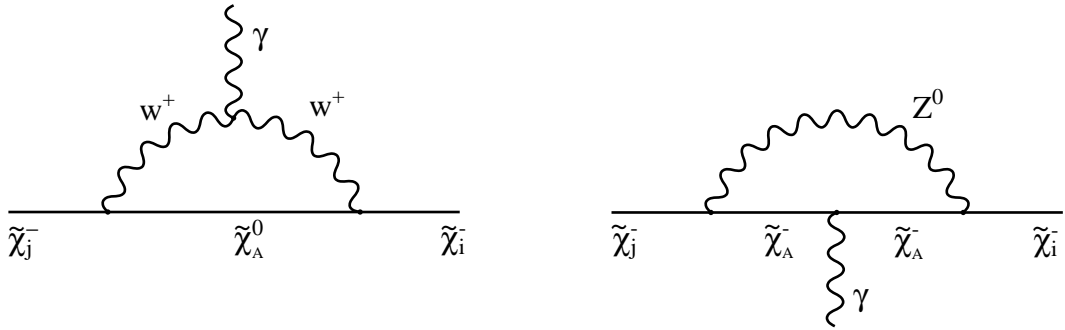}
\caption{Generic Feynman diagrams for 
\protect\( A_{L,Rij}^{G}\protect \).}
\label{fig1}
\end{figure}

\begin{figure}[htbp]
\includegraphics[bb=0 15 440 170]{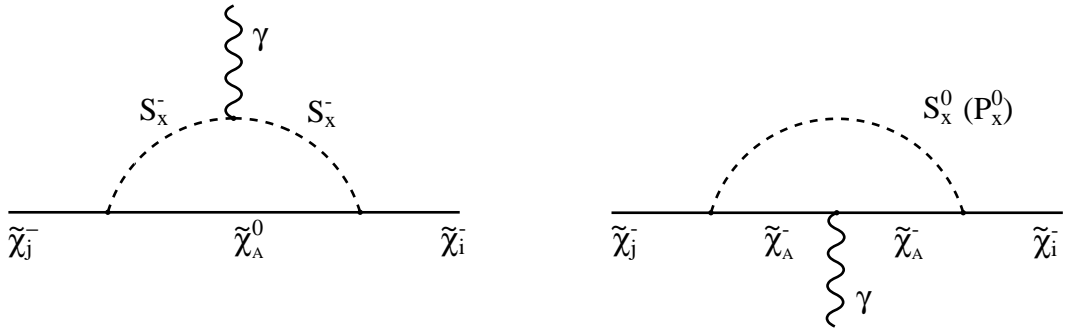}
\caption{Generic Feynman diagrams for $A_{L/R}^{S,ij}$.}
\label{fig2}
\end{figure}

\noindent
The one--loop contributions to  $A_{L,R}$ in the model under
consideration arise from the diagrams of Figs.~\ref{fig1}--\ref{fig3}
\begin{eqnarray}
\label{ALAR}
A_{Lij}&=&A^{S}_{Lij}+A_{Lij}^{G}+A_{Lij}^{Q}\\
A_{Rij}&=&A^{S}_{Rij}+A_{Rij}^{G}+A_{Rij}^{Q}
\end{eqnarray}
The partial contributions on the above expression correspond to the
addition of the sets of diagrams represented in each figure:
\begin{eqnarray}
\label{eq:pam2}
A_{L,Rij}^{G}&=&A_{L,Rij}^{N^{0}-W^{\pm }}+A_{L,Rij}^{C^{\pm
}-Z^{0}}\\
\label{eq:pam1}
A^{S}_{L,Rij}&=&A_{L,Rij}^{N^{0}-S^{\pm }}+ A_{L,Rij}^{C^{\pm
}-S^{0}}+A_{L,Rij}^{C^{\pm }-P^{0}}\\
\label{eq:pam3}
A_{L,Rij}^{Q}&=&A_{L,Rij}^{d\gamma -\widetilde{u}}+
A_{L,Rij}^{d-\widetilde{u}\gamma }+ A_{L,Rij}^{u\gamma
-\widetilde{d}}+A_{L,Rij}^{u-\widetilde{d}\gamma }
\end{eqnarray}
The superscripts in each contribution on the right denote the fermion
and boson internal lines of the corresponding diagram.  For the
quarks-squarks diagrams we include the symbol $\gamma$ to indicate
whether the photon is attached to the fermion or the boson line.  We
follow the notation of \cite{Romao:1999up} indicating by $S^{\pm }$ the
eigenstates of the charged scalar mass matrix, by $S^{0}$ and $P^{0}$
the eigenstates for the sneutrino--Higgs scalar mass matrices,
CP--even and CP--odd, respectively.

The contributions to \( A_{L,Rij}^{G} \) arise from the diagrams in
Fig.~\ref{fig1}.  The index \( A=1,\ldots,5\) corresponds to the
eigenstates of the chargino--lepton mass matrix, while the indices \(
i,j=1,2,3 \) correspond to the lepton generation indices in the limit
of the MSSM with R--parity conservation.  These diagrams will become
the SM contribution to \( {\mathcal{L}}_{eff} \), Eq.~(\ref{leffg2}),
in the limit \( \epsilon _{i}=0 \). In the case of the SM it provides
the main contribution to \( a_{\mu } \), no contribution for charged
LFV processes when neutrinos are considered massless and a very
suppressed contribution for the values of \( m_{\nu _{i}} \)
compatible with the experimental limits~\cite{P..}.

The contributions to $A_{L/R}^{S,ij}$ arise from the three diagrams of
Fig.~\ref{fig2}, where the index \( X \) refers to the eigenstates of scalar
mass matrices. \( S^{\pm } \) are the eigenstates of the \( 8\times 8
\) charged Higgs--slepton mass matrix and \( S^{0},P^{0} \) represent
the eigenstates of the \( 5\times 5 \) neutral Higgs--sneutrino scalar
and pseudoscalar mass matrices, respectively. In the limit where
R--parity is conserved these three diagrams will be combined in the
two supersymmetric diagrams contributing to the \( A_{L/R} \) in the
MSSM. In this this limit, these diagrams are flavor conserving when
the soft-terms are universal as given by minimal Supergravity version
of the MSSM.

$A_{L/R}^{Q,ij}$ arise from the four diagrams of Fig.~\ref{fig3},
where the indices \( X=1,\ldots,6 \) refer to the eigenstates of \(
6\times 6 \) squark mass matrices and indices \( a=1,2,3 \) are the
quark generation indices. These diagrams are not present when
R--parity is conserved.

\subsection{$BR(l_j\rightarrow l_i \gamma)$  for Flavor Violating Processes}

The branching ratio for the rare lepton decays $l_j\rightarrow l_i
\gamma$ is given in the literature~\cite{hisano} and we do not repeat
the derivation here. The result is

\begin{equation}
\label{eq:brm}
BR(l^{-}_{j}\rightarrow l^{-}_{i}\gamma)=\frac{48\pi^3 \alpha }{G_F^2}
\left(\left| A_{Lij}\right| ^{2}+\left| A_{Rij}\right| ^{2}\right)
\end{equation}

\noindent 
where the amplitudes $A_{Lij}$ and $A_{Rij}$ were defined in
Eq.~(\ref{ALAR}). The complete expressions for the amplitudes
corresponding to these processes in the BRpV model are given in the
Appendix~\ref{ap:C}.  In their derivation we have neglected the mass of
the outgoing fermion.

\subsection{The Muon Anomalous Magnetic Moment}

The expression for the muon anomalous magnetic moment can be obtained from
the Lagrangian given in Eq.~(\ref{leffg2}). One obtains \cite{eno},

\begin{equation}
\label{eq:amu}
a^\mu=\frac{(g_\mu-2)}{2}=-m^{2}_{\mu }(A^{\mu\mu}_{L}+A^{\mu\mu}_{R})
\end{equation}
The amplitudes $A^{\mu\mu}_{L/R}$ can also be obtained from the formulas of 
Appendix~C by including the effect of $m_\mu$ in both external lines 
of the diagrams. To do that we just have to include a factor of 
2 in the part of the amplitude containing the function $f_P, \ P=N,C,W,Z$: 
\begin{equation}
A^{a_{\mu }}_{L/R}=A^{22}_{L/R}(f_P(x)\rightarrow 2f_P(x))
\end{equation}

\begin{figure}[htbp]
\includegraphics[bb=0 15 440 165]{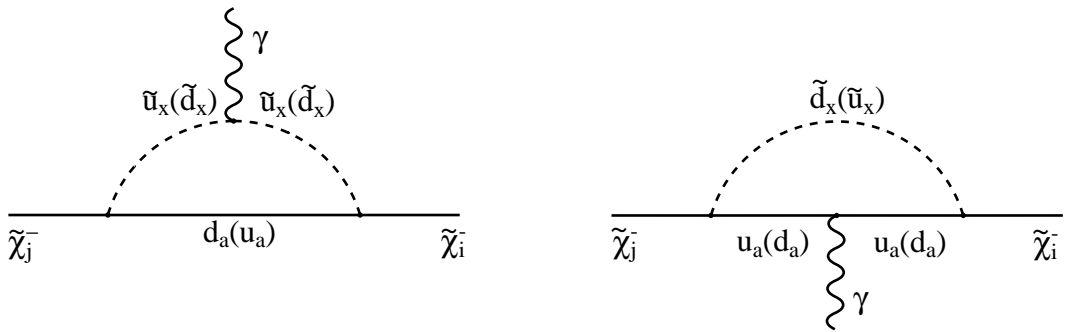}
\caption{Generic Feynman diagrams for $A_{L/R}^{Q,ij}$.}
\label{fig3}
\end{figure}

\section{Results}

\label{sec:Results}

\subsection{The Parameter Space}
\label{sec:parameters}

The BRpV model that we consider adds more free parameters to the ones
already present on the MSSM. However, if we consider the
phenomenological constraints imposed on the MSSM by the limits on the
mass of the lightest neutral $CP$--even Higgs $m_h$, by the
$BR(b\rightarrow s \gamma)$ and by the value of the $a^\mu$, as well
as those derived from neutrino physics on the BRpV parameters, we can
narrow the space of parameters such that generic predictions for $
BR(\mu \rightarrow e \gamma)$ and $ BR(\tau \rightarrow \mu \gamma)$
can be made.

We assume the parameter space of the MSSM with universal soft-terms
and GUT--unification,
\begin{equation}
\alpha _{G},\, \, M_{GUT},\, \, m_{0},\, \, M_{1/2},\, \, 
\tan \beta ,\, \, \mu ,\, \, B,\, \, h_{E},\, \, h_{U},\, \, h_{D},\, \, A_{0}
\end{equation}
with the addition of the BRpV parameters,
\begin{equation}
 \epsilon _{i},\, \, B_{i},\, \, v_{i},\, \, \, \, \, \, i=1,2,3 .
\end{equation}
$A_0$ is defined such that $A_I(GUT)=A_0\cdot h_I, \ I=U,D,E$.
The quantities $\alpha_G=g_G^2/4 \pi$ ($g_G$ being the GUT gauge
coupling constant) and $M_{GUT}$ are evaluated consistently with the
experimental values of $\alpha_{em}$, $\alpha_s$, and $\sin^2\theta_W$
at $m_Z$. We integrate numerically the RGE's for the BRpV model, at
two loops in the gauge and Yukawa couplings and a one loop in the soft
terms, from $M_{GUT}$ down to a common supersymmetric threshold
$M_S\sim\sqrt{m_{\tilde{t}_1}m_{\tilde{t}_2}}$. From this energy to
$m_Z$, the RGE's of the SM are used.

As we explained before, the minimum conditions of the effective scalar
potential allows us to express the values of $\mu ,\, \, B,\, \,
B_{1}\, \, B_{2},\, \, B_{3}$ in terms of \( \tan \beta ,\, \,
\epsilon _{i},\, \, v_{i} \). These are evaluated at the scale
$M_S$. The value of $\mu$ obtained at this scale is similar to the one
obtained by minimizing the effective potential with the complete
1-loop MSSM contributions \cite{dn92}. The 1-loop contributions
arising from Rp-violating terms for these parameters are comparatively
much smaller.

We fix the elements of the quark Yukawa matrices at the GUT scale,
consistently with the experimental values of the quark masses and the
absolute values of the CKM matrix elements. In the case of the charged 
leptons we have to make sure that the three lightest eigenvalues of the
chargino--charged lepton matrix are consistent with the experimental values of
the charged lepton masses.

The values of $m_0$ and $m_{1/2}$ are chosen in region of the
parameter space favored by the considerations presented in
Ref.~\cite{eno}, so that we can compare our results with typical
predictions for the $BR(\mu\rightarrow e \gamma)$ in the MSSM with a
\textit{``see--saw''} mechanism, as discussed in
Ref.~\cite{Carvalho:2001ex}. Obviously, since our model breaks
R-parity, the LSP is not a dark matter candidate and therefore the
cosmological preferred areas of Ref.~\cite{eno} do not apply to our
study. However, the restriction of considering points in the
$m_0-m_{1/2}$ plane such that $m_h>113\ \rm{GeV}$, is the most
restrictive. The SUSY contribution to $a^\mu$ \cite{g-2} favors the sign of
$\mu$ to be positive for the choice of SUSY parameters given below. 
We found that the upper bound of
Eq.~(\ref{eq:g2range}) (see below) on $\delta a^{\mu}$ is less
restrictive than the one imposed by $m_h>113\ \rm{GeV}$.  We analyze
three sets of SUSY parameters,
\begin{itemize}
\item[a)] $\tan\beta=10$, $m_{1/2}=400\ {\rm GeV}$, $m_0=200\ {\rm GeV}$, 
$A_0=0$, $m_\nu=0.1\ {\rm eV}$.
\item[b)] $\tan\beta=30$, $m_{1/2}=400\ {\rm GeV}$, $m_0=300\ {\rm GeV}$, 
$A_0=0$, $m_\nu=0.1\ {\rm eV}$.
\item[c)] $\tan\beta=30$, $m_{1/2}=600\ {\rm GeV}$, $m_0=300\ {\rm GeV}$, 
$A_0=0$, $m_\nu=0.1\ {\rm eV}$.  
\end{itemize}

The six free BRpV parameters \( \epsilon _{i},\, \, v_{i} \) reduce to
three if we take into account the constraints imposed by the
predictions for neutrino oscillations in this model, as given in
Ref.~\cite{Romao:1999up}.  By setting the atmospheric neutrino anomaly
scale to the magnitude of the tree level non--zero value of one of the
neutrinos, Eq.~(\ref{mnutree}), we fix the value \( \sum _{i}
\Lambda_{i}^2 \) for each SUSY point, where the $\Lambda _{i}$ were
defined in Eq.~(\ref{lambdai}). We then follow the discussion of
Ref.~\cite{Romao:1999up}, where it was shown that the conditions \(
\Lambda_{3}\simeq\Lambda_{2}\simeq 5\times \Lambda_{1} \) satisfy both
the atmospheric neutrino anomaly mixings and the CHOOZ
result~\cite{chooz}. We then obtain a linear relationship between each
couple \( \epsilon _{i},\, \, v_{i} \).

Therefore we study the dependence of the process under consideration
on the values of \( \epsilon_{1},\, \, \epsilon_{2},\, \, \epsilon_{3}
\) for a neutrino mass of $m_{\nu_3} = 0.1\ \rm{eV}$, on the upper limit of
the allowed range for the atmospheric neutrino anomaly. For comparison
purposes we also present some results for $m_{\nu_3}=0.05\, \rm{eV}$
on the middle of that range, anf for $m_{\nu_3}=1\, \rm{eV}$.

We will assign random values to $\epsilon_1$ and $\epsilon_2$ 
in the range:
\begin{equation}
-2\times 10^{-3}\ \hbox{GeV} 
\geq  \epsilon_1, \epsilon_2 \geq -60\ \hbox{GeV}.
\end{equation}
However the requirement that $m_{\nu_2} <0.01\ \rm{eV}$ will exclude
values of $|\epsilon_i| > 0.1\ \rm{GeV}$. This will be explicitly
shown in our results.  The value of $\epsilon_3$ is kept fixed since
our results are not altered when it varies on the above range.

\subsection{The Branching Ratio for \protect\( l^{-}_{j}\rightarrow l^{-}_{i}
\gamma \protect \) }

\begin{figure}[htb]
\begin{center}
\includegraphics[height=10cm,bb=22 28 591 516]{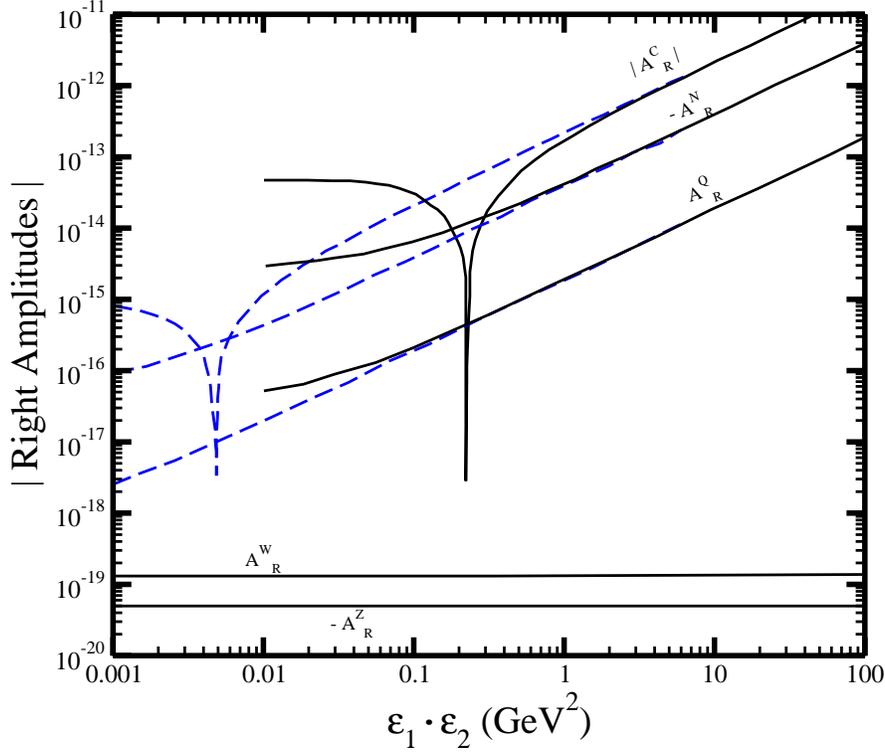}
\end{center}
\caption{Partial Amplitudes $A_R$ corresponding to the choice of 
parameters b) of section 6.1 for $\epsilon_1=-5$~GeV (solid) and 
$\epsilon_1=-0.1$~GeV (dash). The values of $A_R^C$ change sign on each 
branch of the curve, the left one corresponds to positive values. }
\label{fig4}
\end{figure}

We perform a full numerical analysis with the exact diagonalization of
matrices involved in the computation of branching ratios. The main 
contribution for $BR(\mu\rightarrow e \gamma)$, Eq.~(\ref{eq:brm}), comes 
from the amplitudes
$A_R$. The partial contributions from the various diagrams listed in
Eq.~(\ref{eq:pam2}--\ref{eq:pam3}) are displayed in Fig.~\ref{fig4},
for the set of parameters b).  We have found that they are all
independent of $\epsilon_3$ and that they display a linear behavior as
a function of the product $\epsilon_1 \cdot \epsilon_2$, when this
product is larger than 0.1 GeV except for the cancellation observed in
$A_{R}^{C}=A_{R}^{C^{\pm }-S^{0}}+A_{R}^{C^{\pm }-P^{0}}$. The values
of the amplitudes arising from the diagrams of Fig.~\ref{fig1} depend
on the $\epsilon$'s through the $\Lambda$'s which are kept fixed, and
therefore remain constant.  We also found the contributions of the
diagrams of Fig.~\ref{fig3} to be of the same order of magnitude. As
we can see, the sum of all of them, $A_{R}^{Q}$, is almost a linear
function of $\epsilon_1 \cdot \epsilon_2$.  The amplitudes arising
from the diagrams of Fig.~\ref{fig2} are the dominant ones. The one
mediated by the neutralino ($A^N_R$) is smaller than the dominant
chargino exchange ($A^C_R$), except in the range where the cancellation takes
place.

\begin{figure}[htbp]
  \begin{center}
    \begin{picture}(115,100)
      \put(0,0){\includegraphics[height=10cm]{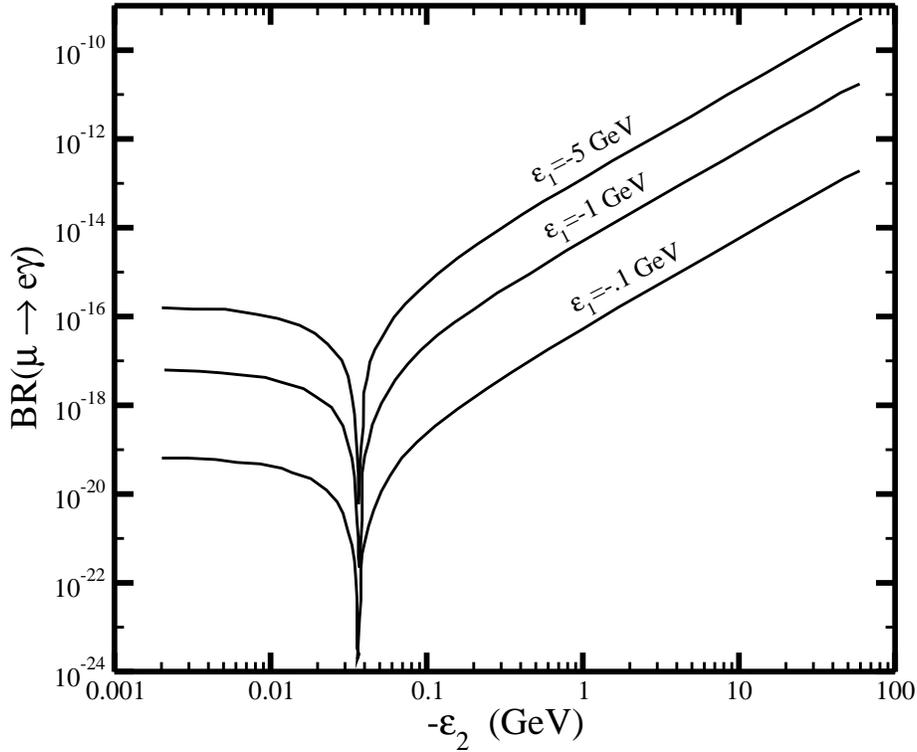}}
    \end{picture}
    \caption{BR(\( \mu \rightarrow e\gamma  \)) vs. $-\epsilon_2$ for 
             $\epsilon_1=-0.1,-1,-5$ GeV for case b.}
    \label{fig5}
  \end{center}
\end{figure}

\begin{figure}
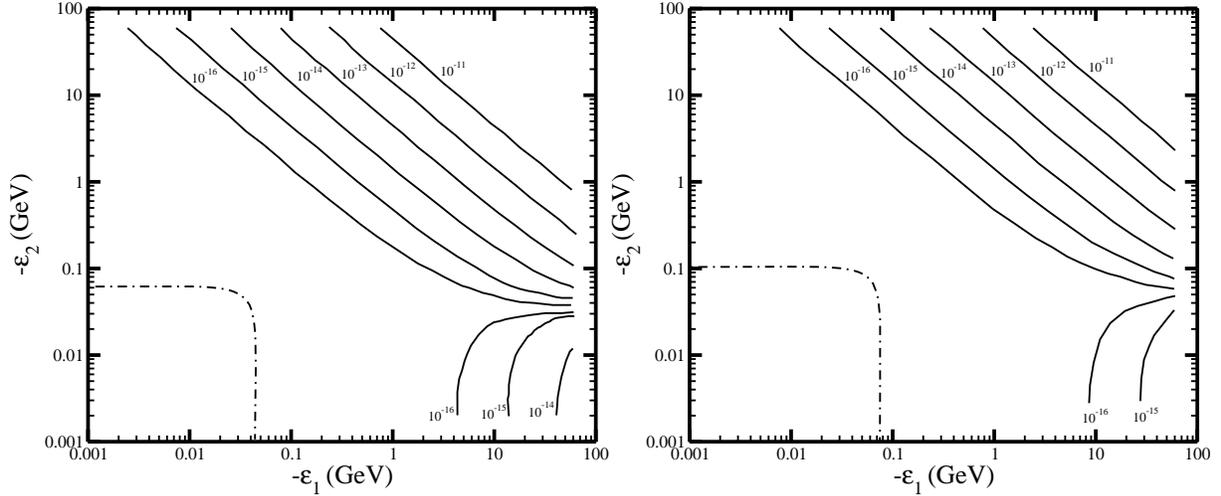

    \begin{picture}(160,69)
      \put(1,0){\includegraphics[width=80mm]{areab1.eps}}
      \put(81,0){\includegraphics[width=80mm]{areac1.eps}}
    \end{picture}
    \caption{Contour plot for BR($ \mu \rightarrow e\gamma$) in 
            $ \epsilon _{1}$--$\epsilon _{2}$ plane for case b (left) 
and c (right). The dash lines correspond to $m_{\nu_2}=0.01$~eV.}
    \label{fig6}
\end{figure}

\begin{figure}
  \begin{center}
    \begin{picture}(115,100)
      \put(0,0){\includegraphics[height=10cm]{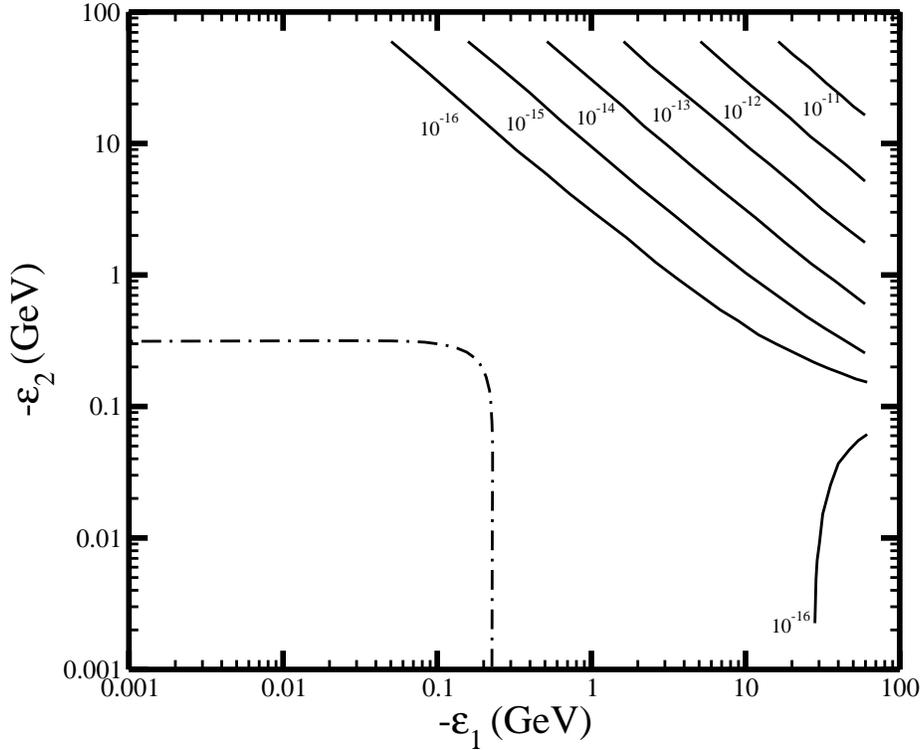}}
    \end{picture}
    \caption{Same as Fig.~\ref{fig6} for case a.}
    \label{fig7}
  \end{center}
\end{figure}

\begin{figure}
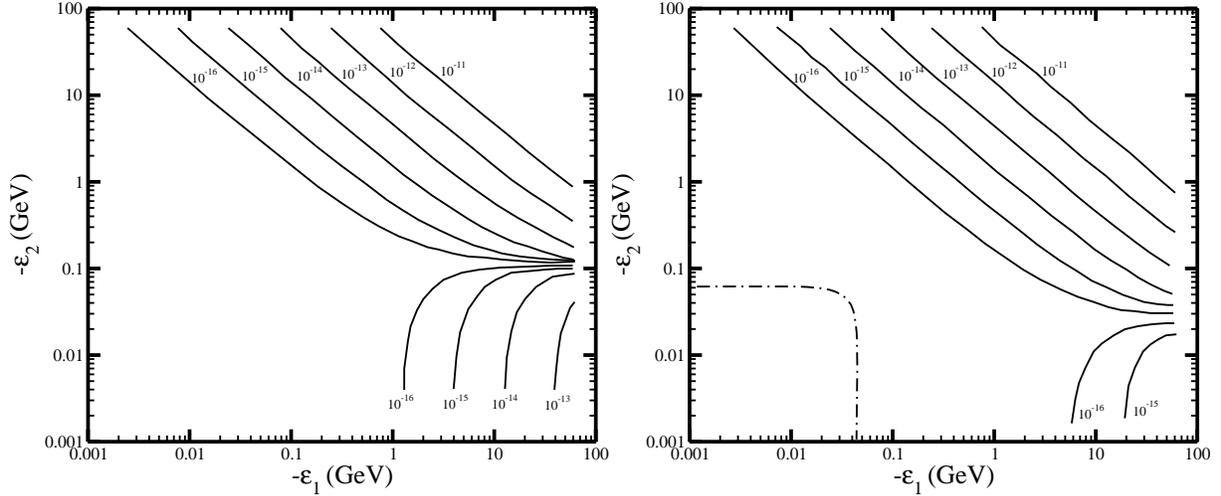

  \begin{center}
    \begin{picture}(160,69)
      \put(1,0){\includegraphics[width=80mm]{B1eV1.eps}}
      \put(81,0){\includegraphics[width=80mm]{B005eV1.eps}}
    \end{picture}
    \caption{Same as Fig.~\ref{fig6}, case b for $m_{\nu_3}=1$~eV (left) and 
 $m_{\nu_3}=0.05$~eV (right).}
    \label{fig8}
  \end{center}
\end{figure}

We can also observe in Fig.~\ref{fig4} that the cancellation which
appear in the $A_{R}^{C}$ depends on the value of
$\epsilon_1$. The values that we show correspond to two different
choices of $\epsilon_1$ (note that if we allow $\epsilon_1$ to change
randomly, as we have done with the other amplitudes, the values for
$A_{R}^{C}$ would be a distribution of dots). The behavior
of $A_{R}^{C}$ cannot be attributed to an accidental
cancellation between the scalar and pseudoscalar parts as one may
naively expect, on the contrary both parts add constructively and
almost vanish simultaneously. The behavior of that amplitude can be
explained when we identify which are the particles running in the
loops of Fig.~\ref{fig2} that are responsible for the main
contributions: $X=1, \ A=1$ and $X=4, \ A=1,2$. Then we can obtain an
accurate approximation for the amplitude by using the formulas given in
the Appendices~\ref{ap:A},~\ref{ap:B}~and~\ref{ap:C}.  Let's consider
$A_{R}^{C^{\pm }-S^{0}}$ since the contribution of the corresponding
pseudoscalar exchange is almost identical. We get from Eq.~(\ref{eq:ALCS})
for the dominant contributions,
\begin{equation}
A_{R34}^{C^{\pm }-S^{0}}\!\! \approx\! -\frac{1}{32\pi ^{2}}\! \left[
\frac{1}{m_{S^{0}_{1}}^{2}}h_{C}(x_{11})\frac{m_{\chi _{1}^{\pm }}}
{m_{\mu}}V_{R311}^{ccs}V_{L411}^{ccs*}\!+\!
\frac{1}{m_{S^{0}_{4}}^{2}}\sum_{A=1}^2 
h_{C}(x_{A4})\frac{m_{\chi _{A}^{\pm }}}{m_{\mu}}V_{R3A4}^{ccs} 
V_{L4A4}^{ccs*} 
\right]
\end{equation}
Using the definitions of Appendix~\ref{ap:B} we find,
\begin{equation}
V_{R311}^{ccs}V_{L411}^{ccs*}\approx -\frac{g h_\mu}{\sqrt{2}}\, 
U_{32}U_{14}, \ \ \ \ 
V_{R3A4}^{ccs}V_{L4A4}^{ccs*}\approx \frac{g h_\mu}{\sqrt{2}}\, 
U_{34}V_{A1}U_{A2}
\end{equation}
where $h_\mu$ is the Yukawa coupling of the muon. We can then write,
\begin{equation}
A_{R34}^{C^{\pm }-S^{0}}\approx F_1\  U_{32}U_{14}+ F_2 \  U_{34}
\end{equation}
where
\begin{eqnarray}
  \label{eq:F1F2}
  F_1&=&\frac{g h_\mu}{32 \sqrt{2}\, \pi^{2}}
\frac{1}{m_{S^{0}_{1}}^{2}}h_{C}(x_{11})\frac{m_{\chi _{1}^{\pm }}}
{m_{\mu}}\\
F_2&=&-\frac{g h_\mu}{32 \sqrt{2}\, \pi^{2}}
\frac{1}{m_{S^{0}_{4}}^{2}}\sum_{A=1}^2 
h_{C}(x_{A4})\frac{m_{\chi _{A}^{\pm }}}{m_{\mu}} V_{A1}U_{A2}
\end{eqnarray}
The quantities $F_1$ and $F_2$ are independent of the BRpV parameters
$\epsilon_i$ and can be evaluated given the SUSY parameters.  The
dependence of the amplitude $A_{R34}^{C^{\pm}-S^{0}}$ on the
$\epsilon_i$ comes from the matrix elements $U_{32}$, $U_{14}$ and
$U_{34}$.  Using the Appendix~\ref{ap:A} we can find approximate
expressions for these matrix elements that display explicitly this
dependence. We get,
\begin{equation}
 U_{34}\approx - \frac{\epsilon_1 \epsilon_2}{\mu^2}+
\frac{\epsilon_1 \Lambda_2}{v_d \mu^2}, \ \ \ 
U_{32}\approx \frac{\epsilon_1}{\mu}, \ \ \ 
U_{14}\approx - U^L_{12} \frac{\epsilon_2}{\mu}.  
\end{equation}
Hence we can find the value of $\epsilon_2$ at which $A^C_R \simeq 0$,
\begin{equation}
\label{cancellation}
\epsilon_2\approx \frac{\Lambda_2 F_2 /v_d}{F_2+F_1 U^L_{12}}
\end{equation}
The position of the cancellation changes, with the value of the SUSY
parameters and also with value of $\Lambda_2$, as we can see from
Eq.~(\ref{cancellation}).  This explains the qualitative changes we
find in Figs.~\ref{fig5}--\ref{fig8}. 

Some of the the amplitudes contributing to the BR($\mu\rightarrow e
\gamma$) presented above have been previously discussed in
Refs.~\cite{frank,Cheung:2001sb}. We agree with
Ref.~\cite{Cheung:2001sb} in that the main contribution arises from
$A_R^C$ except for the values of parameters affected by the
cancellation mentioned above. However we find smaller values for
$A_R^Q$ than the ones quoted in \cite{frank}.

The contribution of $A_L$ to the branching ratio is negligible 
compared with $A_R$, due to the fact that the matrix 
$U$ is replaced by $V$ with suppressed mixings. This holds even for 
the element $V_{34}$. As we can see in Appendix~\ref{ap:A} this element is 
determined by $V_R$ which is obtained in a similar way as $V_L$ for the 
matrix U. However, we observe that the main contribution to $V_{34}$ 
is suppressed by a factor $m_e/m_\mu$  with respect to the corresponding 
one in $U_{34}$. 

Fig.~\ref{fig5} shows the impact of the cancellation in $A_R^C$ on
the predictions of BR($\mu \rightarrow e\gamma$) for the choice of 
parameters b).  As we can deduce from Eq.~(\ref{cancellation}) the
value of the $\epsilon_2$ at which the cancellation in $A_R^C$ takes
place depend on the values of the SUSY parameters. This determines 
the shape of the curves of constant BR in 
Figs.~\ref{fig6},~\ref{fig7}~and~\ref{fig8}. Since the main 
contribution comes from the
chargino mediated diagram of Fig.~\ref{fig2}, we can expect that the
$BR$ increases with $\tan\beta$ and decreases as $m_{1/2}$ grows.  The
increase of $m_0$ produces the same qualitative effect as the increase
of $m_{1/2}$, however it has a lower impact on BR than the changes
in $m_{1/2}$.
 
Our results can be compared with the predictions of a model based in
the MSSM with a {\it ``see--saw''} mechanism presented in
Ref.~\cite{Carvalho:2001ex}, where the results for $BR(\mu \rightarrow
e \gamma)$ are of order $10^{-13}$ for case a) and between $10^{-12}$
and $10^{-13}$ for b) and c). If we observe our predictions for case
b) on the left graph of Fig.~\ref{fig6}, we can see that the model
predicts ratios of $10^{-11}-10^{-13}$ for values of $|\epsilon_1|$
and $|\epsilon_2|$ ranging from $1$ to $10$~GeV (independently of the
value of $\epsilon_3$). Values in the range of $0.1$ to $1$~ GeV would 
lead to rates of order $10^{-14}-10^{-16}$, still interesting for the 
next generation experiments \cite{psi,prism}. Similarly, a window of
$0.1 < -\epsilon_1 ,-\epsilon_2< 1$ GeV, is crossed only by the
$10^{-16}$ line in case c) and by lines below this value for case
a). 
Such values of $|\epsilon_i|$ are however excluded if one takes in
account the constraint coming from the solar neutrinos mass
scale. This is shown in Figs.~\ref{fig6} and \ref{fig7}, where the
dashed line gives the upper limit on the $|\epsilon_i|$ as obtained
from Eq.~(\ref{eq:mneu2}) for the requirement that $m_{\nu_2} <
0.01\ \rm{eV}$ .

In these figures we can appreciate that the parameters that enhance
the ratios also make $m_{\nu_2}$ larger. From Eq.(\ref{eq:mneu2}) we
can observe that $m_{\nu_2}$ increases with $\tan\beta$ (through the
dependence on $h_b$) and decreases as the $\mu$-term increases (i.e
with $m_{1/2}$ and $m_0$).  Furthermore, since we have choosen the
$\Lambda$-parameters to be related, $m_{\nu_2}$ is proportional to a
combination of $\epsilon_i^2$ with no accidental cancellation amoung
them. Therefore we can not find suitable values for $\tan\beta$,
$m_{1/2}$ and $m_0$, such that we can find an overlaping of areas with
$m_{\nu_2}<0.01$~eV and BR($\mu\rightarrow e \gamma$) $>10^{-16}$.

\noindent
 
In Fig.~\ref{fig8} we consider $m_{\nu_3}=0.05 \ {\rm eV}$ and
$m_{\nu_3}=1 \ {\rm eV}$ both for case b).  This decreases (increases)
the values of the $\Lambda$'s by about a factor of $\sqrt{2}$
($\sqrt{10}$), respectively. By looking at the parameterization of the
matrix U in Appendix A we can infer that these changes in the
$\Lambda$'s have not a decisive impact on the $\mu \rightarrow e
\gamma$ rates.  The reason for this is that the dominant contributions
to $A_{R34}^{C^{\pm }-S^{0},P^{0}}$ are determined by the matrix
$\Omega_L$ and its elements that depend explicitly on the $\epsilon$'s
are much larger than the ones containing $\Lambda$'s (at least for
values of $ \epsilon$'s leading to relevant ratios). However the
position of the cancellation on $A_{R34}^{C^{\pm }-S^{0}}$ depends on
$\Lambda_2$ as we can see in Eq.~(\ref{cancellation}) and it therefore
determines the changes in the figures.

The changes on the $\Lambda$'s have only a direct impact on the
smaller contributions, such as the ones arising from the diagrams of
Fig.~\ref{fig1} and on the $A_L$, which size is controlled by the
elements of $\Omega_R$, which, as we have said, are several orders of
magnitude below the main contribution coming from
$\Omega_L$.

The predictions for $BR(\tau \rightarrow \mu \gamma)$ that we obtain
with this model are of the same order as those for $BR(\mu \rightarrow
e \gamma)$, the reason being that we have assumed the $\Lambda$'s to
be of the same order of magnitude, as it is required to explain
neutrino oscillations.  The results in this case are independent of $
\epsilon_1$. If we consider values for $ \epsilon_2$ and $ \epsilon_3$
in the same range as in Fig.~\ref{fig6} we obtain similar 
curves. This result contrasts with the LFV results on the framework of
the R-parity conserving MSSM, where $BR(\mu \rightarrow e \gamma)$ is
typically suppressed by several orders of magnitude with respect to 
$BR(\tau \rightarrow \mu \gamma)$. In this case the hierarchy of Yukawas
couplings makes a distinction between the two processes.

We conclude therefore that the values of the parameters of the BRpV
model that successfully explain the solar and atmospheric neutrino
data~\cite{Romao:1999up}, predict rates for $\mu \rightarrow e\gamma$,
$\tau \rightarrow \mu \gamma$ and $\tau \rightarrow e \gamma$, that
are well below the current limits and well as those of planned
experiments.

\subsection{The Muon Anomalous Magnetic Moment}

The difference on the value of the muon anomalous magnetic moment
found in the BNL E821 measurement \cite{Brown:2001mg} with respect to
the SM prediction, which originally was considered to be $2.6\,
\sigma$ is now reduced to $1.6\, \sigma$ after a theoretical error has
been corrected \cite{cog2}. When the $2\, \sigma$ range is considered,
the allowed values for contributions beyond the SM become,
\begin{equation}
\label{eq:g2range}
-6\times 10^{-10}\leq \delta a_\mu \equiv  
a_\mu^{exp}-a_\mu^{SM}\leq 58\times 10^{-10},
\end{equation}
Several studies \cite{eno} indicate that the MSSM extension of the SM 
can account for this discrepancy. When R--parity is broken the SUSY 
particles are allowed to enter in the SM diagrams (Fig.~\ref{fig1}) and 
conversely the SM particles run in the SUSY loops
(Figs.~\ref{fig2}~and~\ref{fig3}).

The contribution due to the R-parity violating operators to $\delta
a_\mu$ is obtained by subtracting from the amplitudes arising from
Fig.~\ref{fig1} ($\delta a^Z_\mu(RpV)$, $\delta a^W_\mu(RpV)$) and
Fig.~\ref{fig2} ($\delta a^{\tilde{\chi}^{\pm}}_\mu(RpV)$,
$\delta a^{\tilde{\chi}^0}_\mu(RpV)$) the contribution of the SM and
the MSSM, respectively (which are obtained in the limit of vanishing
$\epsilon_i$'s). The contribution from Fig.~\ref{fig3} ($a^q_\mu$) is
not present in the MSSM.  All these contributions are found to be
small when the R-parity violating terms are associated with neutrino
masses of experimental interest.

In Table~\ref{tab:table1} we show the different contributions to
$\delta a_\mu$ for the selected values of $m_0, m_{1/2}$ and
$\tan\beta$ discussed in the section~\ref{sec:parameters}.  The main
contributions to $\delta a_\mu$ arise basically from the MSSM
components of the diagrams in Fig.~\ref{fig2}
($a^{\tilde{\chi}^+}_\mu$, $a^{\tilde{\chi}^0}_\mu$).  The
contribution from BRpV operators just adds a small percentage to the
total values arising from physics beyond the SM. The values that we
show correspond to the maximum value obtained in the conditions for
the $\epsilon$--parameters described in section~\ref{sec:parameters},
when we allow $|\epsilon_1|,\ |\epsilon_2|$ to range 
from $0$ to $60~\rm{GeV}$ (the
result is almost independent of the value of $\epsilon_3$).

\begin{table} 
\begin{center}
\begin{tabular}{|c|c|c|c|c|c|c|c|}\hline
\ \ &
\ \ $a^{\tilde{\chi}^+}_\mu$ \ \ &
\ \ $a^{\tilde{\chi}^0}_\mu$ \ \ &
\ \ $a^q_\mu$ \ \ &
$\delta a^{\tilde{\chi}^+}_\mu(RpV)$&
$\delta a^{\tilde{\chi}^0}_\mu(RpV)$&
$\delta a^Z_\mu(RpV)$&
$\delta a^W_\mu(RpV)$ \\\hline
a&$9.6$&$0.45$&$-1.5\cdot 10^{-4}$&$7.5\cdot 10^{-2}$ &$6.1\cdot
10^{-3}$&$-0.25$&$0.51$\\ \hline 
b&$25.8$&$-1.1$&$-1.5\cdot 10^{-3}$&$0.35$&$2.5\cdot
10^{-2}$&$-0.28$&$0.52$\\ \hline 
c&$13.8$&$0.19$&$-4.1\cdot 10^{-4}$&$0.11$&$1.9\cdot 10^{-2}$ &$-0.14$
&$0.28$\\ \hline 
\end{tabular}
\caption{Contributions to $a_{\mu}$ from the graphs in
  Fig.~\ref{fig1}--\ref{fig3} (in units of $10^{-10}$). Cases a--c, 
refer to the choice of  parameters given in section~\ref{sec:parameters}.  See
  section~\ref{sec:Expressions} for details on the notation.} 
\label{tab:table1}
\end{center}
\end{table}

\section{Conclusions}

\label{sec:Conclusions}
We studied the LFV in one--loop induced rare processes 
$l_j\rightarrow l_i \gamma, \ i\neq j $ in SUSY models with bi--linear 
R--parity violation. In this context, the R--parity violating interactions 
can explain the neutrino masses and mixings without adding 
new fields to the particle content of the MSSM which represent an 
appealing alternative to the {\it ''see--saw''} mechanism. In this work 
we addressed the question of whether neutrino masses in the sub--eV range 
can be compatible with rates for charged LFV processes of experimental 
interest. 

We have performed an exhaustive study of the interactions and of the
SUSY parameters involved in the processes $l_j\rightarrow l_i \gamma,
\ i\neq j $ and contributing to the muon anomalous magnetic
moment. For the case of rare decays we find the diagram mediated by
the Higgs--sneutrino scalars to be the dominant. Contributions arising
from diagrams including lepton--squark and lepton--gauge bosons
vertices, possible in this model, are very suppressed in the range of
neutrino masses considered in this work. Regarding $a_\mu$, the
additional contributions introduced by the R--parity violating
interactions modify in a small percentage the value obtained in the
MSSM limit.

As in a previous analysis \cite{Cheung:2001sb} we find the BR($\mu
\rightarrow e \gamma$) to be very sensitive to the product $\epsilon_1
\cdot \epsilon_2$. However the presence of a cancellation in the main
amplitude contributing to this process (which we have analyzed in
detail through an accurate parameterization of the matrices $U,V$),
makes our contour plots sensitive to the values of $\epsilon_1$ and
$\epsilon_2$ in most of our examples. The rate increases with
$\tan\beta$ as it is the case in the MSSM with {\it ''see--saw''}
mechanism.  However as the one--loop induced neutrino masses will also
grow with $\tan\beta$, the requirement that $\Delta m^2_{12}$ is
compatible with the solar neutrino data excludes the region in
parameter space where the BR($\mu \rightarrow e \gamma$) could be of
experimental interest. On the other hand the rates for $\tau
\rightarrow \mu \gamma$ found in our study are of the same order as
the ones for $\mu \rightarrow e \gamma$, therefore also out of the
experimental range.

Unlike the situation in these models, the rates for $\tau
\rightarrow \mu \gamma$ found in our study are of the same order as
the ones for $\mu \rightarrow e \gamma$, therefore out of the
experimental range.

To conclude, we must say, that the obtained results for the $\mu
\rightarrow e \gamma$ show us that if the BRpV model is the
explanation for both the solar and atmospheric neutrino ocillations,
the predicted LFV will not be testable at PSI~\cite{psi} or at
PRISM~\cite{prism}. The correlations of the BRpV parameters with the
neutralino decays, as proposed in Ref.~\cite{Porod:2000hv}, will
remain the main test of the model.

{\bf ACKNOWLEDGMENTS}
This project was supported in part by the TMR Network of the EC under
contract HPRN-CT-2000-00148.  D. F. C.  and M. E. G. acknowledge
support from the `Funda\c c\~ao para a Ci\^encia e Tecnologia' 
under contracts PRAXIS XXI/
BD/9416/96 and SFRH/BPD/5711/2001 respectively.  We are
grateful to P. Nogueira for his help with the figures presented in this
work.

\newpage 

\appendix
\section{Chargino--Charged Lepton Mass Matrix}
\label{ap:A}

The chargino--charged lepton  mass matrix, in the basis of of
Eq.~(\ref{eq:basisch}), takes the form: 
\begin{equation}
M_{C}=\left[ \begin{array}{cc}
M_{\chi^\pm} & E^\prime\\
E & M_E
\end{array}\right] 
\end{equation}
where $M_E=\frac{1}{\sqrt{2}}v_{d}h_{E}$ is the charged leptons mass 
matrix and $M_{\chi^\pm}$ is the usual MSSM chargino mass matrix,
\begin{equation}
M_{\chi^\pm}=\left[ \begin{array}{cc}
M_{2} & \frac{1}{\sqrt{2}}gv_{u}\\
\frac{1}{\sqrt{2}}gv_{d} & \mu 
\end{array}\right] 
\end{equation}
The sub-matrix  \( E \) is 
\begin{equation}
E=\left[ \begin{array}{cc}
\frac{1}{\sqrt{2}}gv_{1} & -\epsilon _{1}\\
\frac{1}{\sqrt{2}}gv_{2} & -\epsilon _{2}\\
\frac{1}{\sqrt{2}}gv_{3} & -\epsilon _{3}
\end{array}\right] 
\end{equation}
and \( E' \) can be written as \( E'=-v.h_{E} \), where \( v \) is defined 
as:
\begin{equation}
v=\left[ \begin{array}{ccc}
0 & 0 & 0\\
\frac{v_{1}}{\sqrt{2}} & \frac{v_{2}}{\sqrt{2}} & \frac{v_{3}}{\sqrt{2}}
\end{array}\right] 
\end{equation}

As the R--parity breaking parameters are small compared with the SUSY
scale, it is possible to have an approximate diagonalization of \( M_{C} \).
This will be very useful in understanding the numerical results as we can
have approximate analytical formulas. This approximate diagonalization
is obtained by using the following parameterization, introduced in
Refs.~\cite{original_ref,dbd}, for \( U^{*} \)and \( V^{\dagger } \) 
\begin{eqnarray}
U^{*}&=&\left[ \begin{array}{cc}
U_{L}^{*} & 0\\
0 & V_{L}
\end{array}\right] \left[ \begin{array}{cc}
1-\frac{1}{2}\Omega ^{\dagger }_{L}\Omega _{L} & \Omega ^{\dagger }_{L}\\
-\Omega _{L} & 1-\frac{1}{2}\Omega _{L}\Omega ^{\dagger }_{L}
\end{array}\right] \\
\vb{24}
V^{\dagger }&=&\left[ \begin{array}{cc}
1-\frac{1}{2}\Omega ^{\dagger }_{R}\Omega _{R} & -\Omega ^{\dagger }_{R}\\
\Omega _{R} & 1-\frac{1}{2}\Omega _{R}\Omega ^{\dagger }_{R}
\end{array}\right] \left[ \begin{array}{cc}
U_{R}^{\dagger } & 0\\
0 & V^{\dagger }_{R}
\end{array}\right] 
\end{eqnarray}
where $U_{L,R}$ are the MSSM rotation matrices,
\begin{equation}
  \label{eq:rotMSSM}
  U^{*}_{L}M_{\chi^\pm}U^{\dagger}_{R}=M_{\chi^\pm}^{diag}\\  
\end{equation}
The matrices $\Omega_{L,R}$ and $V_{L,R}$ are to be determined from
the unitarity of $U$ and $V$, and from the defining condition
\begin{equation}
  \label{eq:rotDef}
  U^* M_C V^{\dagger}=\left[
    \begin{array}{cc}
      M^{diag}_{\chi^\pm}& 0 \cr   
    \vb{20}
    0& M_E^{diag}
    \end{array}
    \right]
\end{equation}

In the literature~\cite{original_ref,dbd}, the matrices $\Omega_{L,R}$
and $V_{L,R}$ were obtained in the approximation $E'=0$. However we
discovered that this approximation was not good enough to explain our
numerical results. So we re-derived the expressions for these matrices
without neglecting $E'$. We get the following expressions for
$\Omega_{L,R}$, 
\begin{equation}
\Omega _{L}=E M_{\chi^\pm}^{-1}=\frac{1}{\det(M_{\chi^\pm})} 
\left[ \begin{array}{cc}
\frac{g}{\sqrt{2}}\Lambda _{1} & -\frac{1}{\mu }(\epsilon _{1} 
\det(M_{\chi^\pm}) +
\frac{1}{2}g^{2}v_{u}\Lambda _{1})\\
\vb{20}
\frac{g}{\sqrt{2}}\Lambda _{2} & -\frac{1}{\mu }(\epsilon _{2}
\det(M_{\chi^\pm}) +
\frac{1}{2}g^{2}v_{u}\Lambda _{2})\\
\vb{20}
\frac{g}{\sqrt{2}}\Lambda _{3} & -\frac{1}{\mu }(\epsilon _{3} 
\det(M_{\chi^\pm}) +
\frac{1}{2}g^{2}v_{u}\Lambda _{3})
\end{array}\right] 
\end{equation}
\begin{eqnarray}
\Omega _{R}&=&\left(E'^{\dagger } + M^{\dagger}_{e} E
  M_{\chi^\pm}^{-1}\right)(M_{\chi^\pm}^{-1})^{T}=  
h^{\dagger }_{E}\left(-v^{\dagger }+
\frac{1}{\sqrt{2}}v_{d}\Omega _{L}\right) 
(M_{\chi^\pm}^{-1})^{T} \nonumber\\
\vb{20}
&=&\frac{1}{\det(M_{\chi^\pm})}\left[ \begin{array}{cc}
\frac{g}{\sqrt{2}}(M_{E})_{i1}\Lambda _{i} & 
-\frac{M_{2}}{v_{d}}(M_{E})_{i1}\Lambda _{i}\\
\vb{20}
\frac{g}{\sqrt{2}}(M_{E})_{i2}\Lambda _{i} & 
-\frac{M_{2}}{v_{d}}(M_{E})_{i2}\Lambda _{i}\\
\vb{20}
\frac{g}{\sqrt{2}}(M_{E})_{i3}\Lambda _{i} & 
-\frac{M_{2}}{v_{d}}(M_{E})_{i3}\Lambda _{i}
\end{array}\right] \cdot (M_{\chi^\pm}^{-1})^{T}
\end{eqnarray}

\noindent
where summation over \( i=1,2,3 \) is implied in each matrix
element. The expression for $\Omega_L$ coincides with the one found in
the literature but the expression for $\Omega_R$ it is different. 
For $V_{L,R}$ we found that instead of the
relation~\cite{original_ref,dbd} 
\begin{eqnarray}
V_{L}M_{E}V^{\dagger}_{R}&=&M_{E}^{diag}
\end{eqnarray}
they should satisfy,
\begin{eqnarray}
V_{L}\left(M_{E} - \Omega_L  E' - \half \Omega_L \Omega_L^T  M_E 
\right)V^{\dagger}_{R}&=&M_{E}^{diag}
\end{eqnarray}
For a general form of the matrix $M_E$ it will be difficult to have an
explicit form for $V_{L,R}$. However for the case, that we consider,
where the matrix $M_E$ is diagonal, we can obtain an analytical approximate
expression for these matrices,
\begin{equation}
  \label{eq:VLR}
  V_{L,R}\simeq\left[
    \begin{array}{ccc}
      1 & \eta^{L,R}_{12} & \eta^{L,R}_{13}\\
      \vb{20}
       -\eta^{*\, L,R}_{12}& 1 & \eta^{L,R}_{23}\\
      \vb{20}
      -\eta^{*\, L,R}_{13} & -\eta^{*\, L,R}_{23}& 1
    \end{array}
    \right]
\end{equation}
where
\begin{eqnarray}
  \label{eq:etaL}
  \eta_{ij}^L&=& \frac{\epsilon_i\epsilon_j}{2\, \mu^2}\,
  \frac{m_i^2+m_j^2}{m_i^2-m_j^2} +
  \frac{\epsilon_i\Lambda_j}{ \mu^2 v_d}\left[
  \frac{g^2 v_u v_d}{2\, \det(M_{\chi^\pm})} - 
  \frac{m_j^2}{m_i^2-m_j^2} \right] \nn \\
  \vb{20}
  &&+
  \frac{\epsilon_j\Lambda_i}{ \mu^2 v_d}\left[
  -\frac{g^2 v_u v_d}{2\, \det(M_{\chi^\pm})} - 
  \frac{m_i^2}{m_i^2-m_j^2} \right]
  -
  \frac{\Lambda_i\Lambda_j}{ \det(M_{\chi^\pm})\, \mu^2}\,
  \frac{g^2 v_u}{v_d}\,  \frac{m_i^2+m_j^2}{m_i^2-m_j^2} \\
\vb{24}
  \label{eq:etaR}
  \eta_{ij}^R&=& \frac{\epsilon_i\epsilon_j}{ \mu^2}\,
  \frac{m_i m_j}{m_i^2-m_j^2} -
  \frac{\epsilon_i\Lambda_j}{ \mu^2 v_d}
  \frac{m_i m_j}{m_i^2-m_j^2} 
  -\frac{\epsilon_j\Lambda_i}{ \mu^2 v_d}
  \frac{m_i m_j}{m_i^2-m_j^2}\nn \\
  \vb{20}
  &&
  -\frac{\Lambda_i\Lambda_j}{\det(M_{\chi^\pm})\, \mu^2}\,
  \frac{2\, g^2 v_u}{v_d}\,  \frac{m_i m_j}{m_i^2-m_j^2} 
\end{eqnarray}
and $m_i$ are the charged lepton physical masses. 

Putting everything together we can find analytical expressions for the
matrix $U$ that will be useful in explaining our results. We get,
\begin{eqnarray}
  \label{eq:uaprox}
  U^*_{2+i,1}&\simeq&-\frac{g}{\sqrt{2}}\,
  \frac{\Lambda_i}{\det(M_{\chi^\pm})}\\
  \vb{20}
  U^*_{2+i,2}&\simeq&-\Omega_{Li2}\simeq \frac{\epsilon_i}{\mu}\\
  \vb{20}
  U^*_{1,2+i}&\simeq&U^*_{L12}\, \Omega_{Li2}\simeq -U^*_{L12}\,
  \frac{\epsilon_i}{\mu} \\
  \vb{20}
  U^*_{2,2+i}&\simeq&U^*_{L22}\, \Omega_{Li2}\simeq -U^*_{L22}\,
  \frac{\epsilon_i}{\mu} \\
  \vb{20}
  U^*_{2+i,2+j}&\simeq&\left(V_L -\half \Omega_L \Omega_L^T\right)_{ij}
\end{eqnarray}
For further reference we give the approximate expression for
$U^*_{3,4}$. We get
\begin{eqnarray}
  \label{eq:u34}
  U^*_{34}&\simeq&-\frac{\epsilon_1\epsilon_2}{\mu^2}
  +\frac{\epsilon_1 \Lambda_2}{\mu^2 v_d} -\frac{g\, v_u \epsilon_2
  \Lambda_1}{\det(M_{\chi^\pm}) \mu^2}\\
  \vb{20}
  &\simeq&-\frac{\epsilon_1\epsilon_2}{\mu^2}
  +\frac{\epsilon_1 \Lambda_2}{\mu^2 v_d} 
\end{eqnarray}
where we have  assumed that the parameters are in the ranges described in
section~\ref{sec:parameters}.

\section{The Couplings}
\label{ap:B}

The relevant part of the Lagrangian, using four component spinor notation is,
\begin{eqnarray}
  \label{eq:lag}
  {\mathcal{L}}&=&\left[\vb{16}
\overline{\chi _{i}^{-}}(V^{cns}_{LiAX}P_{L}+
V^{cns}_{RiAX}P_{R})\chi _{A}^{0}S^{-}_{X}+
\overline{\chi _{i}^{-}}(V^{cd\widetilde{u}}_{LiAX}P_{L}+
V^{cd\widetilde{u}}_{RiAX}P_{R})d_{A}\widetilde{u}^{*}_{X} \right. \nn \\
&&\left.\vb{16}
+\overline{\chi _{i}^{+}}(V^{cu\widetilde{d}}_{LiAX}P_{L}
+V^{cu\widetilde{d}}_{RiAX}P_{R})u_{A}\widetilde{d}^{*}_{X}
+\overline{\chi _{i}^{-}}\gamma ^{\mu }(V^{cnW}_{LiA}P_{L}
+V^{cnW}_{RiA}P_{R})\chi _{A}^{0}W_{\mu }^{-}+h.c. \right] \nn \\ 
&&\vb{20}
+\overline{\chi_{i}^{-}}(V^{ccs}_{LiAX}P_{L}
+V^{ccs}_{RiAX}P_{R})\chi_{A}^{-}S^{0}_{X}  
+\overline{\chi_{i}^{-}}(V^{ccp}_{LiAX}P_{L}
+V^{ccp}_{RiAX}P_{R})\chi _{A}^{-}P^{0}_{X} \nn \\  
&&\vb{20}
+\overline{\chi _{i}^{-}}\gamma ^{\mu }(V^{ccZ}_{LiA}P_{L}
+V^{ccZ}_{RiA}P_{R})\chi _{A}^{-}Z_{\mu }^{0} 
\end{eqnarray}
The definition of these couplings is given in the following
sections. These definitions extend those of Ref.~\cite{Romao:1999up} which
conventions we follow.

\subsection{Chargino-Neutralino-Charged Scalars}

\begin{eqnarray}
\label{eq:cnsL}
V^{cns}_{LiAX}&\hskip -4mm=\hskip -4mm&
-\rho _{A}\!\left[\vb{16}
gR^{S^{\pm}}_{X2}\!\left(\frac{1}{\sqrt{2}}N^{*}_{A2}V^{*}_{i2} +\!
N^{*}_{A4}V^{*}_{i1}\right)\!+\!g'\!\left(\frac{1}{\sqrt{2}}R^{S^{\pm}}_{X2}
N^{*}_{A1}V^{*}_{i2}+\!\sqrt{2}R^{S^{\pm }}_{X5+\alpha}
V^{*}_{i2+\alpha}N^{*}_{A1}\right)\right. \nn \\   
&&\left. \hskip 7mm \vb{16}
+R^{S^{\pm }}_{X2+\alpha }h^{\alpha \beta }_{E}V^{*}_{i2+\beta }
N^{*}_{A3} - R^{S^{\pm }}_{X1}N^{*}_{A4+\alpha } h^{\alpha \beta}_{E}
V^{*}_{i2+\beta }\right] \\
\vb{22}
V^{cns}_{RiAX}&\hskip -4mm=\hskip -4mm&
\eta _{i}\!\left[\vb{16}
gR^{S^{\pm}}_{X1}\left(\frac{1}{\sqrt{2}}N_{A2}U_{i2}-N_{A3}U_{i1}\right) +
gR^{S^{\pm}}_{X2+\alpha } \left(\frac{1}{\sqrt{2}} U_{i2+\alpha }N_{A2} -
N_{A4+\alpha}U_{i1}\right) \right. \nn \\ 
&&\left. \vb{16}
+\!\frac{g'}{\sqrt{2}}\left(R^{S^{\pm }}_{X1}N_{A1}U_{i2} \!+\!
R^{S^{\pm}}_{X2+\alpha}U_{i2+\alpha } N_{A1}\right) \!\!+\!
\left(U_{i2}N_{A4+\alpha} \!-\! N_{A3}U_{i2+\alpha}\right) h^{\alpha \beta }_{E}R^{S^{\pm }}_{X5+\beta}
\right]
\end{eqnarray}
where the indices have the following ranges: $A=1,\ldots,7$,
$i=1,\ldots,5$, $\alpha,\beta=1,\ldots,3$ and $\rho_A$ ($\eta_i$) are
the signs of the neutralinos (respectively charginos) as they are
obtained from the numerical evaluation of the eigenvalues~\cite{Romao:1999up}.

\subsection{Chargino-Chargino-CP Even Neutral Scalars}
\begin{eqnarray}
  \label{eq:ccsL}
V^{ccs}_{LiAX}&=&-\eta_{A}\frac{1}{\sqrt{2}} \left[ \vb{16}
g\left( R^{S^{0}}_{X1}U^{*}_{A2}V^{*}_{i1} +
  R^{S^{0}}_{X2}U^{*}_{A1}V^{*}_{i2}+R^{S^{0}}_{X2+\alpha }
  U^{*}_{A2+\alpha }V^{*}_{i1}\right) \right. \nn \\
\vb{20}
&& \left. \vb{16}\hskip 15mm
+\left( R^{S^{0}}_{X1}U^{*}_{A2+\alpha } -
  U^{*}_{A2}R^{S^{0}}_{X2+\alpha }\right) h^{\alpha \beta
}_{E}V^{*}_{i2+\beta }\right] \\
\vb{20}
 V^{ccs}_{RiAX}&=&V^{ccs*}_{LAiX} 
\end{eqnarray}
\subsection{Chargino-Chargino-CP Odd Neutral Scalars}
\begin{eqnarray}
  \label{eq:ccpL}
V^{ccp}_{LiAX}&=&
i\eta _{A}\frac{1}{\sqrt{2}} \left[\vb{16}
g\left(R^{P^{0}}_{X1}U^{*}_{A2}V^{*}_{i1} +
  R^{P^{0}}_{X2}U^{*}_{A1}V^{*}_{i2} +
  R^{P^{0}}_{X2+\alpha}U^{*}_{A2+\alpha }V^{*}_{i1}\right) \right.\nn \\ 
\vb{20}
&& \left. \vb{16} \hskip 15mm
+\left(U^{*}_{A2}R^{P^{0}}_{X2+\alpha } -
  R^{S^{0}}_{X1}U^{*}_{A2+\alpha}\right) h^{\alpha
  \beta}_{E}V^{*}_{i2+\beta }\right] \\
\vb{20}
V^{ccp}_{RiAX}&=&V^{ccp*}_{LAiX} 
\end{eqnarray}
\subsection{Chargino-Neutralino-\protect\( W^{\pm }\protect \)}
\begin{eqnarray}
  \label{eq:cnWL}
 V^{cnW}_{LiA}&=&-\eta _{i}\, \rho _{A}\, g\left[
N^{*}_{A2}U_{i1}+\frac{1}{\sqrt{2}}\left(N^{*}_{A3}U_{i2} +
  N^{*}_{A4+\alpha}U_{i2+\alpha}\right) \right] \\
\vb{20}
V^{cnW}_{RiA}&=&g\left[
\frac{1}{\sqrt{2}}N_{A4}V^{*}_{i2} -N_{A2}V^{*}_{i1}\right]
\end{eqnarray}
\subsection{Chargino-Chargino-\protect\( Z^{0}\protect \) }
\begin{eqnarray}
  \label{eq:ccZL}
  V^{ccZ}_{LiA}&=&\eta _{i}\,\eta _{A}\, \frac{g}{\cos \theta
    _{W}}\left[ \frac{1}{2}U_{i1}U^{*}_{A1} + \left(\frac{1}{2}-\sin
    ^{2}\theta_{W}\right) \delta _{iA}\right] \\
\vb{20}
V^{ccZ}_{RiA}&=&\frac{g}{\cos \theta_{W}}\left[V^{*}_{i1}V_{A1} +
    \frac{1}{2}V^{*}_{i2}V_{A2} -\sin ^{2}\theta _{W} \delta _{iA}\right] 
\end{eqnarray}
\subsection{Chargino-Quark Down-Squark Up}
\begin{eqnarray}
  \label{eq:cduL}
 V^{cd\widetilde{u}}_{LiAX}&=&\eta ^{d}_{A}\left[
-gV^{*}_{i1}R^{\widetilde{u}}_{X\alpha }R^{d*}_{LA\alpha }+V^{*}_{i2}R^{d*}_{LA\alpha }h^{\alpha \beta }_{U}R^{\widetilde{u}}_{X3+\beta }\right] \\
\vb{20}
V^{cd\widetilde{u}}_{RiAX}&=&\eta _{i}\left[
U_{i2}R^{\widetilde{u}}_{X\alpha } h^{\alpha \beta}_{D}R^{d}_{RA\beta} 
\right]
\end{eqnarray}
\subsection{Chargino-Quark Up-Squark Down}

\begin{eqnarray}
  \label{eq:cudL}
  V^{cu\widetilde{d}}_{LiAX}&=&\eta _{i}\eta ^{u}_{A}\left[
-gU^{*}_{i1}R^{\widetilde{d}}_{X\alpha }R^{u*}_{LA\alpha } +
U^{*}_{i2}R^{u*}_{LA\alpha} h^{\alpha
  \beta}_{D}R^{\widetilde{d}}_{X3+\beta} \right] \\
\vb{20}
 V^{cu\widetilde{d}}_{RiAX}&=&V_{i2}R^{\widetilde{d}}_{X\alpha }
 h^{\alpha \beta}_{U}R^{u*}_{RA\beta } 
\end{eqnarray}

\section{Amplitudes}
\label{ap:C}

We collect here the various amplitudes corresponding to the diagrams
of Figs.~\ref{fig1}--\ref{fig3}. In these amplitudes the mass of the
outgoing fermion was neglected. We give only the amplitudes $A_L$
because the $A_R$ can be obtained from these with the substitution
rule
\begin{equation}
A_{Rij}= A_{Lij}(L/R\rightarrow R/L).
\end{equation}

\subsection{Neutralinos--Charged Scalars}
\begin{equation}
\label{eq:AmpNS+-}
A_{Lij}^{N^{0}-S^{\pm }}=\sum ^{5}_{A=1}\sum ^{8}_{X=1}
\frac{1}{32\pi^{2}} \frac{1}{m_{S^{\pm }_{X}}^{2}}
\left[f_{N}(x_{AX})V_{LiAX}^{cns}V_{LjAX}^{cns*} +
  h_{N}(x_{AX})\frac{m_{\chi_{A}^{0}}}{m_{l_{j}}} V_{LiAX}^{cns}
  V_{RjAX}^{cns*} \right]
\end{equation}
with \( x_{AX}= \left(\frac{m_{\chi _{A}^{0}}}{m_{S^{\pm}_{X}}}
\right)^{2} \) and the functions \( f_{N},h_{N} \) given by
\begin{equation}
\label{eq:fN}
f_{N}(x)=\frac{1-6x+3x^{2}+2x^{3}-6x^{2}\ln x}{6(1-x)^{4}}\hskip 10mm 
h_{N}(x)=\frac{1-x^{2}+2x\ln x}{(1-x)^{3}}
\end{equation}

\subsection{Charginos--CP Even Neutral Scalars}
\begin{equation}
\label{eq:ALCS}
A_{Lij}^{C^{\pm }-S^{0}} = \sum ^{5}_{A=1}\sum ^{5}_{X=1} -
\frac{1}{32\pi ^{2}}\frac{1}{m_{S^{0}_{X}}^{2}}
\left[f_{C}(x_{AX})V_{LiAX}^{ccs}V_{LjAX}^{ccs*} +
  h_{C}(x_{AX})\frac{m_{\chi _{A}^{\pm }}}{m_{l_{j}}}
  V_{LiAX}^{ccs}V_{RjAX}^{ccs*}\right] 
\end{equation}
with \( x_{AX}=\left(\frac{m_{\chi _{A}^{\pm}}}{m_{S^{0}_{X}}}
\right)^{2} \) and the functions \( f_{C},h_{C} \) given by
\begin{equation}
\label{eq:fC}
f_{C}(x)=\frac{2+3x-6x^{2}+x^{3}+6x\ln x}{6(1-x)^{4}}\hskip 10mm 
h_{C}(x)=\frac{-3+4x-x^{2}-2\ln x}{(1-x)^{3}}
\end{equation}

\subsection{Charginos--CP Odd Neutral Scalars}
\begin{equation}
\label{eq:ALCP}
A_{Lij}^{C^{\pm }-P^{0}} = \sum ^{5}_{A=1}\sum ^{5}_{X=1} -
\frac{1}{32\pi ^{2}}\frac{1}{m_{P^{0}_{X}}^{2}}
\left[f_{C}(x_{AX})V_{LiAX}^{ccp}V_{LjAX}^{ccp*} +
  h_{C}(x_{AX})\frac{m_{\chi_{A}^{\pm }}}{m_{l_{j}}}
  V_{LiAX}^{ccp}V_{RjAX}^{ccp*}\right] 
\end{equation}
with \( x_{AX}=\left(\frac{m_{\chi _{A}^{\pm }}}{m_{P^{0}_{X}}}\right)^{2} \).

\subsection{Quarks--Squarks}
\begin{equation}
\label{eq:ALdgsu}
\hskip -5pt
A_{Lij}^{d\gamma -\widetilde{u}} =\! \sum ^{3}_{A=1}\sum
^{6}_{X=1}3(-\frac{1}{3})\frac{1}{32\pi^{2}}
\frac{1}{m_{\widetilde{u}_{X}}^{2}}  \left[
  f_{C}(x_{AX})V_{LiAX}^{cd\widetilde{u}} V_{LjAX}^{cd\widetilde{u}*}
  \!+\! h_{C}(x_{AX})\frac{m_{d_{A}}}{m_{l_{j}}}
  V_{LiAX}^{cd\widetilde{u}}V_{RjAX}^{cd\widetilde{u}*} \right]
\end{equation}
with \( x_{AX}=\left(\frac{m_{d_{A}}}{m_{\widetilde{u}_{X}}}\right)^{2} \).
\begin{equation}
\label{eq:ALugsd}
\hskip -5pt
A_{Lij}^{u\gamma -\widetilde{d}} = \sum^{3}_{A=1}\sum^{6}_{X=1} 3
(\frac{2}{3})\frac{1}{32\pi^{2}}  \frac{1}{m_{\widetilde{d}_{X}}^{2}}
\left[f_{C}(x_{AX})V_{RiAX}^{cu\widetilde{d}}
  V_{RjAX}^{cu\widetilde{d}*} +
  h_{C}(x_{AX})\frac{m_{u_{A}}}{m_{l_{j}}}V_{RiAX}^{cu\widetilde{d}}
  V_{LjAX}^{cu\widetilde{d}*} \right]
\end{equation}
with \( x_{AX}=\left(\frac{m_{u_{A}}}{m_{\widetilde{d}_{X}}}\right)^{2} \).
\begin{equation}
\label{eq:ALdsug}
\hskip -5pt
A_{Lij}^{d-\widetilde{u}\gamma } = \sum
^{3}_{A=1}\sum^{6}_{X=1}3(\frac{2}{3})\frac{1}{32\pi^{2}}
\frac{1}{m_{\widetilde{u}_{X}}^{2}}
\left[f_{N}(x_{AX})V_{LiAX}^{cd\widetilde{u}}V_{LjAX}^{cd\widetilde{u}*}
  + h_{N}(x_{AX})\frac{m_{d_{A}}}{m_{l_{j}}}
  V_{LiAX}^{cd\widetilde{u}}V_{RjAX}^{cd\widetilde{u}*} \right]
\end{equation}
with \( x_{AX}=\left(\frac{m_{d_{A}}}{m_{\widetilde{u}_{X}}}\right)^{2} \).
\begin{equation}
\label{eq:ALusdg}
\hskip -5pt
A_{Lij}^{u-\widetilde{d}\gamma } \!=\! \sum^{3}_{A=1}
\sum^{6}_{X=1}3(-\frac{1}{3})\frac{1}{32\pi^{2}}
\frac{1}{m_{\widetilde{d}_{X}}^{2}}\left[f_{N}(x_{AX})
  V_{RiAX}^{cu\widetilde{d}}V_{RjAX}^{cu\widetilde{d}*} \! +\!
  h_{N}(x_{AX})\frac{m_{u_{A}}}{m_{l_{j}}}
  V_{RiAX}^{cu\widetilde{d}}V_{LjAX}^{cu\widetilde{d}*} \right]
\end{equation}
with \( x_{AX}=\left(\frac{m_{u_{A}}}{m_{\widetilde{d}_{X}}}\right)^{2} \).

\subsection{W-Neutralinos}
For W--Neutralinos the amplitude, in the unitary gauge (\( \xi
\rightarrow +\infty  \)) is,
\begin{equation}
\label{eq:ALNW}
A_{Lij}^{N^{0}-W^{\pm }} = \sum^{5}_{A=1} -
\frac{1}{32\pi^{2}}\frac{1}{m_{W}^{2}}
\left[f_{W}(x_{A})V_{RiA}^{cnW}V_{RjA}^{cnW*}
 + h_{W}(x_{A})\frac{m_{\chi _{A}^{0}}}{m_{l_{j}}} V_{RiA}^{cnW}
 V_{LjA}^{cnW*}\right] 
\end{equation}
with \( x_{A}=\left(\frac{m_{\chi _{A}^{0}}}{m_{W}}\right)^{2} \) and the
functions \( f_{W},h_{W} \) given by 
\begin{eqnarray}
  \label{eq:fW}
   f_{W}(x)&=&\frac{10-43x+78x^{2} - 49x^{3}+4x^{4}+18x^{3}\ln x} {6(1-x)^{4}}
   \\
\vb{20} 
   h_{W}(x)&=&\frac{-4+15x-12x^{2}+x^{3}+6x^{2}\ln x}{(1-x)^{3}}
\end{eqnarray}

\subsection{Z--Charginos}
For Z--Charginos the amplitude, in the unitary gauge 
(\( \xi \rightarrow +\infty  \)) is,
\begin{equation}
\label{eq:ALCZ}
A_{Lij}^{C^{\pm }-Z^{0}} = \sum^{5}_{A=1}\frac{1}{32\pi^{2}}
\frac{1}{m_{Z}^{2}}\left[f_{Z}(x_{A}) V_{RiA}^{ccZ}V_{RjA}^{ccZ*} +
  h_{Z}(x_{A}) \frac{m_{\chi _{A}^{\pm }}}{m_{l_{j}}} V_{RiA}^{ccZ}
  V_{LjA}^{ccZ*}\right] 
\end{equation}
with \( x_{A}=\left(\frac{m_{\chi _{A}^{\pm }}}{m_{Z}}\right)^{2} \) 
and the functions \( f_{Z},h_{Z} \) given by 
\begin{eqnarray}
  \label{eq:fZ}
  f_{Z}(x)&=&\frac{8-38x+39x^{2}-14x^{3}+5x^{4}-18x^{2}\ln x}{6(1-x)^{4}} 
  \\
  \vb{20}
  h_{Z}(x)&=&\frac{-4+3x+x^{3}-6x\ln x}{(1-x)^{3}}
\end{eqnarray}

\end{document}